\documentclass[sigconf,authorversion,nonacm]{acmart}

\usepackage{tikz}
\usepackage{xcolor}
\usepackage{subcaption}
\usepackage{comment}
\usepackage{chato-notes}
\usepackage{graphicx}
\usepackage{adjustbox}
\usepackage[ruled,vlined]{algorithm2e}
\usepackage[table]{xcolor}
\usepackage{url}
\tcbuselibrary{breakable}
\tcbuselibrary{theorems}
\tcbuselibrary{skins}

\usepackage{pgfplots}
\usepgfplotslibrary{groupplots}
\pgfplotsset{compat=1.18}

\newcommand{\para}[1]{\paragraph{\textnormal{\textbf{#1}}}} 

\newboolean{showcomments}
\setboolean{showcomments}{true} %
\ifthenelse{\boolean{showcomments}}
{
	\newcommand{\nb}[3]{
		{\colorbox{#2}{\bfseries\sffamily\scriptsize\textcolor{white}{#1}}}
		{\textcolor{#2}{$\blacktriangleright$\textsf\small{#3}$\blacktriangleleft$}}}
	 
}{
	\newcommand{\nb}[3]{}
} 

\usepackage{calrsfs}

\DeclareMathAlphabet{\pazocal}{OMS}{zplm}{m}{n}
\DeclareMathAlphabet{\pazobfcal}{OMS}{cmsy}{b}{n}

\newcommand\mybox[2][]{\tikz[overlay]\node[fill=blue!20,inner sep=2pt, anchor=text, rectangle, rounded corners=1mm,#1] {#2};\phantom{#2}}

\newcommand{\gbox}[1]{\mybox[fill=green!20]{#1}}

\newcommand{\rbox}[1]{\mybox[fill=red!20]{#1}}

\definecolor{DefinitionColor}{RGB}{0, 0, 0}

\newtcbtheorem{definitions}{}%
{enhanced,frame empty,interior empty,colframe=DefinitionColor!50!white,
 coltitle=DefinitionColor!50!black,fonttitle=\bfseries,colbacktitle=DefinitionColor!15!white,
 borderline={0.5mm}{-1mm}{DefinitionColor!15!white},
 borderline={0.5mm}{-1mm}{DefinitionColor!70!white,dashed},
 attach boxed title to top center={yshift=-2mm},
 boxed title style={boxrule=0.2pt},varwidth boxed title}{defo}

\usepackage{tcolorbox}

\usepackage{enumitem} 
\newcommand{\uls}{\begin{itemize}[leftmargin=*]}
\newcommand{\ule}{\end{itemize}}
\newcommand{\ols}{\begin{enumerate}[leftmargin=*]}
\newcommand{\ole}{\end{enumerate}}
\newcommand{\li}{\item}

\usepackage{marvosym}
\usepackage{adjustbox}
\usepackage{amsmath}
\usepackage{booktabs}
\usepackage{multirow}
\usepackage{tabularx}
\usepackage{url}
\usepackage{rotating}

\usepackage[normalem]{ulem}

\usepackage[font=small]{caption}

\AtBeginDocument{%
  }

\begin{document}

\title{Generate to Accelerate: Improved Reranking via LLM-Generated Pivot Documents}
\titlenote{Accepted for publication in the Proceedings of the 35th ACM International Conference on Information and Knowledge Management (CIKM 2026)}

\author[N. Sinhababu]{Nilanjan Sinhababu}
\correspondingauthor
\email{nilanjansb@kgpian.iitkgp.ac.in}
\orcid{0000-0003-3594-4143}
\affiliation{
    \institution{Centre for Computational and Data Sciences\\IIT Kharagpur}
    \country{India}
}

\author[S. Bharati]{Soumedhik Bharati}
\email{soumedhik.bharati@mbzuai.ac.ae}
\orcid{0009-0009-6256-207X}
\affiliation{
    \institution{Department of Natural Language Processing\\Mohamed bin Zayed University of Artificial Intelligence}
    \country{United Arab Emirates}
}

\author[D. Ganguly]{Debasis Ganguly}
\email{debasis.ganguly@glasgow.ac.uk}
\orcid{0000-0003-0050-7138}
\affiliation{
    \institution{School of Computing Science\\University of Glasgow}
    \country{United Kingdom}
}

\author[P. Mitra]{Pabitra Mitra}
\email{pabitra@cse.iitkgp.ac.in}
\orcid{0000-0002-1908-9813}
\affiliation{
    \institution{Department of Computer Science and Engineering\\IIT Kharagpur}
    \country{India}
}

\renewcommand{\shortauthors}{Nilanjan et al.}

\begin{abstract}
Common approaches to reduce the computational overhead of reranking models include identifying a candidate set of documents for reranking or constructing comparison graphs to minimize redundant comparisons. For pointwise rankers, determining a candidate set typically involves estimating a query-dependent cutoff based on the scores of the top-ranked documents. In contrast, comparison graphs for listwise approaches are often derived using heuristics, such as propagating local comparisons within sliding windows in a bottom-up fashion or reducing comparisons via pivot-based strategies in a top-down manner.
In this work, we argue that restricting these processes to existing documents in the collection is unnecessary. Instead, we propose leveraging the generative capabilities of large language models to synthesize a pseudo-relevant document for a given query. We then adapt existing reranking approaches and also propose a novel parallel reranking approach to leverage this LLM-generated document as a pivot. Our experiments demonstrate that using LLM-generated pivots for ranked list truncation reduces the number of pointwise ranker inferences by up to 66\%. In both in-domain and out-of-domain settings, we observe speedups of up to $2.95\times$ in listwise reranking, while maintaining comparable or improved retrieval effectiveness.

\end{abstract}



\maketitle

\section{Introduction}
\begin{figure}[t]
    \centering
    \includegraphics[width=\columnwidth]{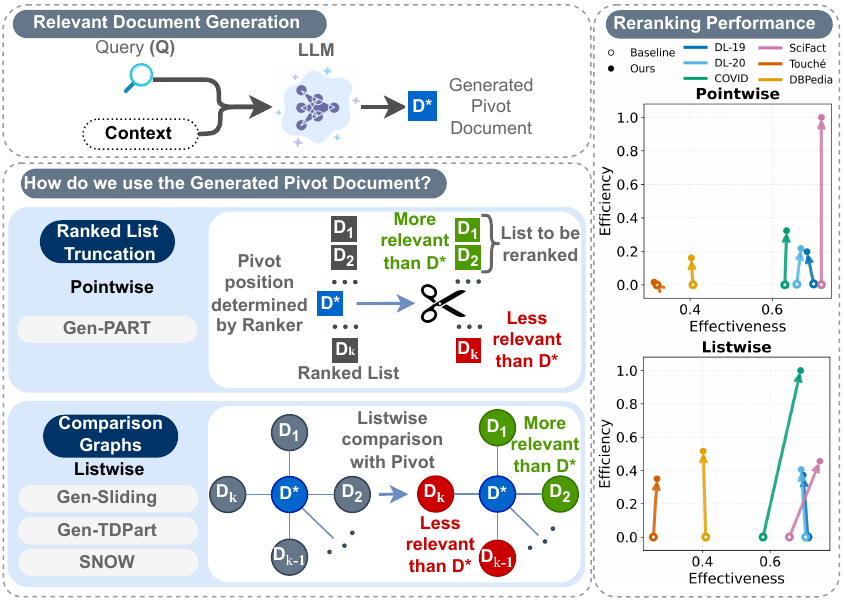}
    \caption{
    Overview of our proposed pointwise and listwise reranking framework. We prompt an LLM to synthetically generate a relevant document ($D^*$) conditioned on $Q$. We then use this pseudo-relevant document as a pivot to determine a cutoff list for pointwise reranking or to determine an accelerated computational workflow for a local document comparison graph in a listwise ranker. Our pivot-based approach improves, or at worst, results in marginal drops in effectiveness (nDCG@10) while significantly improving efficiency (reduced latency) over a range of different test sets.}
    \Description{Overview of our proposed pointwise and listwise reranking framework.}
    \label{fig:overall_method}
\end{figure}

Retrieve-and-rerank pipelines are increasingly employed in modern information retrieval systems. In these pipelines, a \textit{retriever} first generates a top-$k$ set of candidate documents that are likely to be relevant, using either exact or approximate retrieval methods, as commonly implemented in sparse~\cite{robertson1995okapi} and dense indexing approaches~\cite{wang2024textembeddingsweaklysupervisedcontrastive,contriever,gao2022precise}, respectively. The retrieved candidates are then passed to a (re-)\textit{ranker}, which produces a more reliable ordering of the initially retrieved documents~\cite{pradeep2021expandomonoduodesignpatterntext,ma2023fine,zhang-etal-2024-two}.

Although the later-stage rankers in a retrieve-and-rerank pipeline generally provide more accurate relevance estimation, such rankers are typically computationally more expensive than first-stage sparse or dense retrievers~\cite{RetrievalInformation,zhang-etal-2024-two}. Consequently, these systems inherently involve a trade-off between retrieval effectiveness and computational efficiency.
Managing this trade-off requires different techniques---namely, pointwise~\cite{pradeep2021expandomonoduodesignpatterntext} where documents are individually processed or as a list in the listwise~\cite{zhang-etal-2024-two,sun2023chatgpt} approach. The list may consist of subsets of documents retrieved in the first phase, including just a pair of documents in the pairwise approach~\cite{Qin-2024-PRP,Sinhababu-2024-FSPRP}.

To improve the efficiency of pointwise reranking, a widely adopted strategy is to prune the initially retrieved list~\cite{nogueira2019passage} by estimating a \textit{query-dependent cutoff}~\cite{choppycut,meng2024ranked,Lien-Dynamic-RLT-2019}. This assumes that documents beyond the cutoff are unlikely to be relevant. This pruning procedure is commonly referred to as ranked list truncation (RLT)~\cite{where-to-stop,choppycut,Lien-Dynamic-RLT-2019,wu2021learningtruncaterankedlists}.
In contrast, efficiency improvements for listwise (including pairwise) ranking approaches primarily focus on estimating a \textit{sparse comparison graph} for \textit{approximate sorting} (brute-force comparisons are computationally infeasible) that determines which document pairs should be evaluated. These methods typically exploit locality heuristics within the initially retrieved list. For example, prior work~\cite{Qin-2024-PRP,Sinhababu-2024-FSPRP,luo-etal-2024-prp} employs sliding window techniques to restrict comparisons to local neighborhoods, while allowing documents to propagate across overlapping windows in a bottom-up manner, thereby preserving opportunities for broader comparison.

A key limitation of existing approaches for improving reranker efficiency is that they are predominantly either heuristic-based or reliant on shallow information sources. For example, several methods employ heuristic propagation strategies, such as bottom-up~\cite{luo-etal-2024-prp,Qin-2024-PRP,Sinhababu-2024-FSPRP} or top-down~\cite{parry2024topdownpartitioningefficientlistwise} propagation of a subset of top-scoring documents. Furthermore, existing RLT approaches primarily depend on the retrieval scores of top-ranked documents, while largely ignoring the semantic interactions between queries and retrieved documents,  which supervised query performance prediction (QPP) research has shown to be informative~\cite{singh2023unsupervised,datta2022pointwise,arabzadeh2021bert}.

Recent studies suggest that large language models (LLMs) exhibit strong capabilities in capturing semantic relationships and modeling query--document interactions~\cite{thomas2024large,Guglielmo-2023-Perspectives,gao2022precise,wang-etal-2023-query2doc}. Motivated by these findings, we hypothesize that LLMs can be effectively leveraged to improve the efficiency of reranking pipelines in two complementary ways. First, they may help identify a cutoff rank beyond which relevant documents become sparse~\cite{choppycut,Lien-Dynamic-RLT-2019,meng2024ranked}. Second, they may facilitate the construction of comparison graphs that are informed by the semantic content of the top-retrieved documents, rather than relying solely on their positional proximity in the ranked list~\cite{Qin-2024-PRP,luo-etal-2024-prp,Sinhababu-2024-FSPRP,parry2024topdownpartitioningefficientlistwise}.

We propose that such semantic information can be incorporated into reranker-efficiency mechanisms by introducing an explicit query-specific pseudo-relevant document. Such a document which we refer to as a \textit{semantic pivot}, captures the core information need expressed by the query and provides a reference against which retrieved documents can be assessed. In pointwise reranking, the pivot's rank position can define a query-dependent truncation cutoff, assuming that documents scored below it are less likely to be relevant. In listwise and pairwise reranking, the pivot can guide comparison-graph construction by making comparisons depend not only on rank proximity or locality heuristics, but also on each document's semantic relationship to the pivot.

The remaining challenge is how to obtain such a pivot document without requiring human relevance annotations. We address this by leveraging LLMs to generate a synthetic answer or summary document for each query. Recent work has shown that LLMs are effective at modeling query--document relevance and can be used as relevance judges in zero-shot, few-shot, and instruction-tuned settings. These models often achieve strong agreement with human assessors~\cite{Guglielmo-2023-Perspectives}. These findings suggest that LLMs possess sufficient semantic understanding of information needs and relevance criteria to generate a plausible pseudo-relevant document. In our work, the LLM is
therefore not used directly as a judge or reranker; rather, it is used to produce a synthetic document that acts as a semantic pivot for guiding efficiency-oriented decisions in downstream reranking.

\para{Our Contributions}
The primary contributions of this work are summarized as follows:

\ols
    \li \textit{LLM-generated semantic pivot:} We propose using an LLM
    to generate a synthetic answer (generated pivot document in Figure~\ref{fig:overall_method}) for a query.

    \li \textit{Semantic pivot-guided ranked list truncation:} We introduce
    \textbf{Gen-PART} (pointwise method in Figure~\ref{fig:overall_method}), an adaptive RLT method for pointwise ranking.

    \li \textit{Adapting comparison graphs for listwise reranking:} We
    adapt existing listwise comparison heuristics by conditioning document
    comparisons on their position related to the pivot, rather than
    relying solely on document rank positions as shown in Figure~\ref{fig:overall_method}). In particular, we propose \textbf{Gen-Sliding} and \textbf{Gen-TDPart} as adaptations of a standard bottom up \cite{luo-etal-2024-prp,Qin-2024-PRP} and top-down comparison heuristic \cite{parry2024topdownpartitioningefficientlistwise}, respectively.

    \li \textit{Parallel listwise reranking via semantic propagation:}
    We introduce \textbf{SNOW} (Shared Non-Overlapping Window), a pivot-guided listwise comparison graph that enables parallel processing of non-overlapping windows while using the shared pivot to propagate semantic information across windows.
    Code is made publicly 
available for research and 
development\footnote{\url{https://github.com/nilanjansb/DynamicRLT}}.

\ole

\section{Related Work}

\para{LLMs for Judgment and Generation}
LLMs are now extensively used for relevance judgment of documents. 
Several benchmarks of LLMs as judges have been provided by \citeauthor{thomas2024large} and \citeauthor{Guglielmo-2023-Perspectives}. Particularly, UMBRELA~\cite{Upadhyay-2024-UMBRELA} has been proposed as a framework with zero-shot and few-shot prompting. The interplay of LLM judgment, rankers, and nature of documents has been studied by \citeauthor{Balog-LLMEval-2025}. 
Reranking comparison graphs for improved relevance assessment has been proposed by \citeauthor{meng-2026-rerankersrelevancejudges}. \citeauthor{Ruiyang-SelfCalibratedListwiseRerankingLLM-2025} use fine-tuned 
LLMs to generate global scores for the documents in the ranked list, 
improving listwise reranking efficiency. These works demonstrate that LLMs possess strong understanding of query-document relevance.

\para{Ranked List Truncation}
Traditionally, RLT has been formulated as a cascade ranking problem, where the optimal cutoff depth for a ranked list is determined before
reranking to balance effectiveness against efficiency. \citeauthor{Lien-Dynamic-RLT-2019} explored efficient top-$k$ processing by
dynamically pruning candidates unlikely to affect the final evaluation
metric.

For neural retrieval, RLT has evolved from heuristic score-based
thresholding to learned models. \citeauthor{choppycut} introduced
\textit{Choppy}, a transformer-based model that predicts query-specific
cutoffs by analyzing the score distribution of the top-retrieved
documents. Similarly,\citeauthor{wu2021learningtruncaterankedlists}
proposed \textit{AttnCut}, which uses an attention mechanism to model
the dependencies between document scores and the optimal truncation point.
More recently, \citeauthor{meng2024ranked} investigated RLT
specifically for reranking pipelines, demonstrating that fixed-depth
reranking is often suboptimal and that dynamic truncation can significantly
improve the efficiency of LLM-based rerankers. %

\para{LLM-based Reranking}
Recent developments in LLM capabilities and scale, especially large
context windows, have led to the emergence of listwise reranking approaches~\cite{tang_2024_found,gangireddy_2024_first,liu2024slidingwindowsendexploring,sun2023chatgpt}.
Listwise objectives better capture inter-document dependencies than pointwise, pairwise, or traditional cross-encoder
approaches~\cite{sun2023chatgpt,Sinhababu-2025-ModelingRank}. Distilled LLMs, trained via permutation distillation on teacher rankings
from zero-shot prompts, outperform comparable or larger supervised cross-encoders such as
monoT5~\cite{sun2023chatgpt,pradeep2023rankvicunazeroshotlistwisedocument,pradeep2023rankzephyreffectiverobustzeroshot}.

Despite these effectiveness gains, inference efficiency remains a primary 
challenge in deploying listwise LLM rerankers at scale~\cite{parry2024topdownpartitioningefficientlistwise,gangireddy_2024_first}.
Recent supervised approaches, such as FIRST, address this by limiting generation to a single identifier token, thereby improving retrieval
efficiency~\cite{gangireddy_2024_first,chen-enhanceFIRST-2025}. \citeauthor{li2025leveragingreferencedocumentszeroshot} leverage documents 
from the ranked list as reference points for efficient zero-shot LLM 
reranking. \citeauthor{parry2024topdownpartitioningefficientlistwise} proposed
Top-Down Partitioning (TDPart), which replaces sequential window traversal
with a pivot-guided selection that enables parallel reranking of lower-ranked partitions.

\section{LLM based Reference-Document Generation} \label{sec:sec3}
In this section, we first describe the method of generating an LLM response relevant to an input query.
This synthetically generated document (pivot) then determines a dynamic rank cutoff for pointwise rankers (Section \ref{sec:gen-part}) and determines the computation batches for listwise rankers (Section \ref{sec:listwise}).

\para{LLMs-based Relevance Assessments}

Prior work has shown that LLMs can generate relevance judgments for
query-document pairs with high correlation to human
assessors~\cite{Rahmani-2024-LLM4Eval,Upadhyay-2024-UMBRELA,Guglielmo-2023-Perspectives}.
This implies that LLMs possess a graded understanding of query-document
relevance when provided with a relevance scale comparable to that used by
human assessors~\cite{Upadhyay-2024-UMBRELA}. However, existing strategies
use this capability solely for relevance judgment, rather than for
generating documents at a specified relevance grade.
In our work, we hypothesize that if an LLM can
reliably judge that a document belongs to a given relevance level (non-relevant, partially relevant, moderately relevant or highly relevant), it
should similarly be capable of generating a document of that level,
providing a controllable semantic reference for ranking.

\para{Pivot Generation}

\begin{figure}[t]
\centering
\begin{adjustbox}{width=.95\columnwidth}
\begin{tcolorbox}[
    width=\columnwidth,
    colback=white,
    colframe=black,
    arc=3mm,
    boxrule=0.8pt,
    left=4pt, right=4pt, top=4pt, bottom=4pt,
    fonttitle=\small\bfseries
]
\small
You are an expert information retrieval specialist tasked with document generation. Your task is to generate a single document that matches a specific relevance grade and is relevant to the given query.\\
Given the query: ``\texttt{\{Q\}}''\\
Generate a document that would receive a relevance score of ``\texttt{\{$\tau$\}}''\\
\\
Relevance grade score definitions:\\
0: Non-relevant; the document or passage provides no useful information for the query.\\
1: Partially relevant; offers some related information but is of limited utility.\\
2: Somewhat relevant; partially addresses the query but leaves significant information needs unmet.\\
3: Highly relevant; provides an ideal response to the query, often comprehensive and precise.
\end{tcolorbox}
\end{adjustbox}
\caption{Prompt $\pazocal{I}(Q, \tau)$ used to generate the semantic pseudo-relevant pivot document $D^*$ in our proposed reranking workflow. This prompt explicitly instructs an LLM to generate a document matching a target relevance grade $\tau$ for a given query $Q$, where $\tau$ is set to $2$ (somewhat relevant) in our work. The 0--3 grading scale definitions are consistent with the standard guidelines used by human assessors employed by NIST ~\cite{craswell2020overview} and also with the \texttt{UMBRELA} automated relevance evaluation framework~\cite{Upadhyay-2024-UMBRELA}.}
\Description{Pseudo-relevant document generation prompt.}
\label{fig:prompt_pivot}
\end{figure}

The core innovation of our approach is the generation of a pivot document
$D^*$ that encodes query-dependent semantics and enables dynamic truncation
for each query $Q$.
The pivot document $D^*$ is generated from an LLM, it is agnostic
of the first-stage retriever or downstream
reranker, which makes it possible to use it as a reference point of comparison across different reranking models and types (pointwise or listwise).

Since this pseudo-relevant document is intended to serve
as a reference point for comparisons with documents in the initially retrieved
list, we do not generate it at the highest relevance level according to the
standard four-point TREC DL relevance scale, which is also commonly adopted
in LLM-based relevance assessment studies~\cite{Upadhyay-2024-UMBRELA}.
Generating a highly relevant pivot may make the pseudo-relevant document overly
specific to a narrow aspect of the information need. When used to guide the
comparison workflow of a reranker, such a pivot may therefore be too selective,
allowing only a small subset of documents to be retained for subsequent
reranking. Conversely, generating a pivot with a low relevance grade may
produce an overly generic document, causing too many candidates to be retained
and thereby reducing the efficiency gains from truncation.

We therefore
generate the pivot at a moderate relevance level, which is expected to provide a balance between semantic specificity and coverage. In this way, the pivot is
broad enough to capture the main aspects of the information need, while still
selective enough to limit the number of documents passed to the reranker (Figure \ref{fig:prompt_pivot} shows the prompt structure).
Specifically, for our main experiments reported in Section \ref{ss:pointwise-results}, we set this relevance label of the LLM-generated relevant document to 2, i.e., moderately relevant, as per the standard TREC DL relevance scale of $\{0, 1, 2, 3\}$. We also present an ablation of the effect of this relevance label.

More formally speaking, the generation task uses a structured prompt instruction $\pazocal{I}(Q, \tau)$ designed for an instruction-tuned LLM $\psi$ that explicitly conditions the model on both
the query semantics and the target relevance level, as shown in
Figure~\ref{fig:prompt_pivot}. Formally, we prompt an LLM ($\psi$) to
generate a document $D^*(Q;\psi,\tau)$ such that:
\begin{equation}
D^*\equiv D^*(Q;\psi,\tau) = \psi(\pazocal{I}(Q, \tau)),
\label{eq:pivotgen}
\end{equation}
where $\tau$ denotes the target relevance grade, and $D^*$ represents a notation short form of $D^*(Q;\psi,\tau)$.

\begin{figure}[t]
\centering
\begin{tcolorbox}[
    width=0.85\columnwidth,
    colback=white,
    colframe=black,
    arc=3mm,
    boxrule=0.8pt,
    left=4pt, right=4pt, top=4pt, bottom=4pt,
    fonttitle=\small\bfseries
]
\small
\textbf{Query ($Q$):} Do goldfish grow?
\vspace{2pt}
\hrule
\vspace{2pt}
\textbf{Example Document Ranked \underline{Above} the Pivot ($D^+$):} \\
Genetics are the predetermined factor in how \gbox{big your fish will grow}. Goldfish sold as tank suitable will \gbox{grow to about 10 inches} while those sold as pond suitable will \gbox{reach a maximum of about 18 inches.}
\vspace{2pt}
\hrule
\vspace{2pt}
\textbf{Generated Pivot ($D^*$ | $\tau=2$):} \\
\rbox{Setting up a tank} for coldwater fish requires attention to filtration and temperature. Unlike tropical fish, species such as the \rbox{goldfish do not require a heater}, but they do \rbox{produce a significant amount of waste}. This means you will need a robust filter to keep ammonia levels down. It is also important to \rbox{consider the size of your tank} carefully. \gbox{While often sold as small juveniles, these fish will increase} \gbox{in size as they mature}, eventually outgrowing small bowls or cramped environments.
\vspace{2pt}
\hrule
\vspace{2pt}
\textbf{Example Document Ranked \underline{Below} the Pivot ($D^-$):} \\
\rbox{Goldfish food} contains less protein and more carbohydrates than other \rbox{fish food}... Your goldfish need to \rbox{eat proper goldfish food} that meets their specific needs.
\end{tcolorbox}
\caption{This example shows that the document above the pivot is a \textit{highly relevant} document directly addressing the information need of the query.
The document scored below the pivot is a non-relevant document which topically is related to gold fish but fails to address the query. However, our \textit{semantic pivot} ($D^*$) is positioned somewhere in between, i.e., it answers the query (marked in \gbox{green}) along with tangential information (marked in \rbox{red}).}
\label{fig:pivot_comparison}
\Description{Example of pivot document and how it pivots between relevant and non-relevant documents.}
\end{figure}

\section{Generative Pointwise Reranking}
\label{sec:gen-part}

We now describe how to leverage the LLM-generated relevant document (Section \ref{sec:sec3}) for pointwise reranking. In particular, we describe \emph{\textbf{Gen}erative \textbf{PART}itioning} (\textbf{Gen-PART}), a novel RLT approach that leverages generated relevant documents.

\subsection{Pivot Document as a Relevance Threshold}
\begin{figure}[t]
    \centering
    \includegraphics[width=\columnwidth]{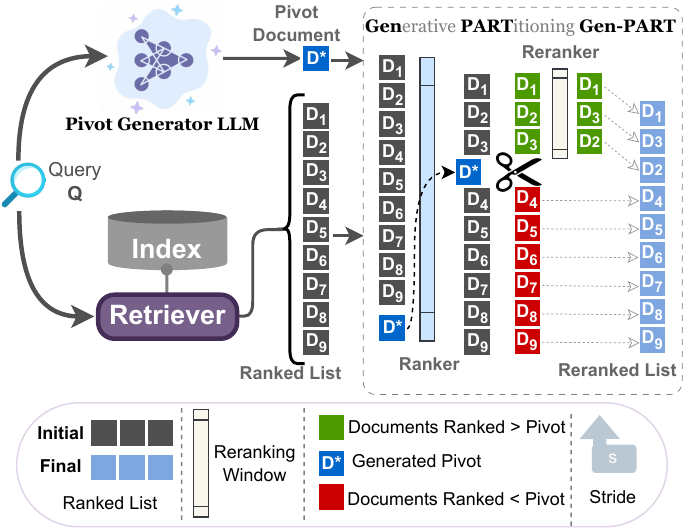}
    \caption{Generation of a reference-document pivot ($D^*$) for a given query
    $Q$ and its corresponding ranked list from a first-stage retriever (e.g., BM25~\cite{robertson1995okapi}). For pointwise setup, Gen-PART uses a single inference to position $D^*$ in the ranked list; all documents scoring above the pivot (green) are forwarded to the reranker.}
    \Description{Workflow diagram showing Gen-PART.}
    \label{fig:Gen-PART}
\end{figure}

First, we use the initial retriever (e.g., BM25) to additionally score the generated pivot document $D^*$ (Equation \ref{eq:pivotgen}).
This requires only a single additional inference, resulting in negligible computational overhead for positioning $D^*$ within the initially ranked list.
As a next step, documents scored higher than the pivot are considered to be the more promising candidates and are forwarded to the reranker for subsequent processing. As per the work in \cite{meng2024ranked}, the remaining documents are appended to the reranked list in the same order as they were initially retrieved, as shown in Figure \ref{fig:Gen-PART}.

The rationale for considering only documents ranked above the pivot for
subsequent processing is based on the assumption that the pivot represents a moderately relevant, sufficiently broad-scoped pseudo-relevant document. Accordingly, documents assigned lower scores than the pivot by the initial retriever are
assumed to be less likely to be highly relevant. Moreover, because the pivot is synthetically generated rather than drawn from the collection, it can potentially capture multiple partially relevant subtopics associated with the query (e.g., Figure \ref{fig:pivot_comparison} shows that the generated document is multi-topical containing information on setting up the ideal water temperature and the tank size for goldfishes to grow). Documents ranked below such a pivot are therefore less likely to contain any of these relevant aspects (e.g., the document ranked below the pivot in Figure \ref{fig:pivot_comparison} is only about goldfish food), providing a basis for treating them as unlikely
candidates for further reranking.

Figure~\ref{fig:pivot_comparison} illustrates this intuition using an example from the TREC DL'19 dataset. In this example, a document ranked above the pivot by the initial BM25 retriever is observed to be highly relevant, with a relevance grade of 3, whereas a document ranked below the pivot is observed to be non-relevant and is effectively removed by the proposed truncation strategy.

Formally, let $L(Q; \theta) = \{D_1,\ldots,D_m\}$ be the list of $m$ top-retrieved documents for a query $Q$ obtained by a retriever model $\theta$. In a retrieve-rerank pipeline, a reranking model $\phi$ is then used to recompute the scores of these documents and obtain a different ordering.
For efficiency reasons, a smaller cutoff $p$ ($p < m$) is selected for this score recomputation with the assumption that beyond a certain rank cutoff, recall starts to saturate, i.e., it is less likely to find more relevant documents, which means that reranking is likely not to be beneficial.
Formally speaking, the cutoff rank $p$ is used to partition the list $L(Q; \theta)$ into
two sub-lists
\begin{equation}
L(D^+, Q;\theta) = \{D_1,\ldots,D_p\},\,\,
L(D^-, Q;\theta) = \{D_{p+1},\ldots,D_m\},
\label{eq:dplus-and-dminus}
\end{equation}
following which the final reranked list is obtained as
\begin{equation}
L(Q; \theta, \phi) = \phi(L(D^+,Q;\theta)) \oplus L(D^-,Q;\theta),
\label{eq:rerankedlist}
\end{equation}
where $\oplus$ is the list concatenation operator.

\subsection{RLT cutoff by Pivot Document's Score}

Our proposed RLT method, \textbf{Gen-PART}, estimates
the cutoff $p$ of Equation \ref{eq:rerankedlist} for each query by modeling relevance more directly as compared to existing RLT approaches \cite{choppycut,Lien-Dynamic-RLT-2019,MtCut,wu2021learningtruncaterankedlists} which primarily rely on information from retrieval-score distributions or other ranked-list statistics.
More precisely speaking, Gen-PART uses an LLM to transform these query-level semantics into a
pivot document.
More formally, we score the generated pivot document $D^*$ using the initial
retriever $\theta$, obtaining the retrieval score $\theta(Q,D^*)$, where $D^*$ is generated according to Equation~\ref{eq:pivotgen}. We then insert $D^*$ into the ranked list at a position $p$ such that the ordering induced by $\theta$ is
preserved:
\begin{equation}
\theta(Q,D_p) \geq \theta(Q,D^*(Q;\psi,\tau)) \geq \theta(Q,D_{p+1}).
\label{eq:scores-insert}
\end{equation}
This value of $p$ then becomes the rank cutoff of Equation \ref{eq:dplus-and-dminus}.
This way of estimating the cutoff makes Equation \ref{eq:rerankedlist} depend not only on the initial retriever $\theta$ and the reranker model $\phi$ but also on the LLM $\psi$ used to generate the pivot document with a relevance level of $\tau$ (Equation \ref{eq:pivotgen}). This means that the rank cutoff, in turn, depends on the information need expressed in the query and also on the content of the documents initially retrieved at different rank positions.

It is worth mentioning that Gen-PART does not require task-specific supervision or parameter training, and can therefore be applied in a zero-shot manner across a wide range of collections and retrieval settings (validated in our experiments in Section \ref{sec:results}).

\section{Generative Listwise Reranking}
\label{sec:listwise}

We now describe how to leverage the LLM-generated relevant document (Section \ref{sec:sec3}) for listwise reranking.

\subsection{Comparison Graphs for Listwise Reranking}

In contrast to a pointwise ranker, which scores each query--document pair
independently, i.e., $\phi: Q \times D \mapsto \mathbb{R}$, a listwise
ranker takes a query and a list of documents as input and returns a reordered
version of that list. Formally, a listwise ranker can be written as
\[
\phi: Q \times \{D_1,\ldots,D_n\}
\mapsto
\{D_{\sigma(1)},\ldots,D_{\sigma(n)}\},
\]
where $\sigma: \mathbb{Z}_n \mapsto \mathbb{Z}_n$ denotes a permutation.
In practice, the number of documents $n$ provided to a listwise ranker is
substantially smaller than the number of initially retrieved documents $m$.
This restriction is necessary because long input lists can both degrade
ranking effectiveness, due to overly large contexts for reliable inference,
and increase inference latency.

Accordingly, efficiency in listwise reranking is not primarily achieved by
truncating the retrieved list, as in pointwise reranking, but by determining
which subsets of documents should be jointly reranked and how information
should propagate across these subsets~\cite{sun2023chatgpt,Qin-2024-PRP,parry2024topdownpartitioningefficientlistwise}.
This process can be viewed as defining a comparison graph over the candidate
documents. Existing approaches typically construct this graph statically,
independently of the query content, and rely on heuristic assumptions about
the initially retrieved ranking. For example, the commonly used sliding-window
strategy~\cite{luo-etal-2024-prp} partitions $L(Q;\theta)$ into overlapping
windows of size $n$, with stride $s$ controlling the degree of overlap. The
listwise ranker is then applied iteratively, usually starting from the bottom
of the initially retrieved list, while propagating the top-$s$ documents from
each window to the next.

More generally, the comparison graph specifies which documents are compared
within each listwise inference call and how the outputs of one call affect
subsequent calls. For the sliding-window strategy, this computation can be
abstractly represented as
\begin{equation}
\phi_s(D_{[m-n+1:m]})
\rightarrow
\phi_s(D_{[m-n-s+1:m-s]})
\cdots
\rightarrow
\phi_s(D_{[1:n]}),
\label{eq:swindow}
\end{equation}
where $D_{[i:j]}$ denotes the sublist $\{D_i,\ldots,D_j\}$ with $j>i$, and
$\phi_s$ denotes the top-$s$ documents selected after reranking a window. The
overlap between consecutive windows enables documents that are promoted by
the ranker to propagate towards earlier positions in the final ranking.

We next propose three methods (Sections \ref{sec:gen-sliding-method}, \ref{sec:gen-tdpart-method} and \ref{sec:snow-method}) that leverage $D^*$, the synthetically generated
pseudo-relevant document defined in Equation~\ref{eq:pivotgen}, to construct
comparison graphs dynamically for listwise reranking. Unlike static
position-based heuristics, these methods use the generated pivot (Sections \ref{sec:sec3} and \ref{sec:gen-part}) to incorporate
query-dependent semantic information into the reranking workflow.

\begin{figure}[t]
    \centering
    \includegraphics[width=0.7\columnwidth]{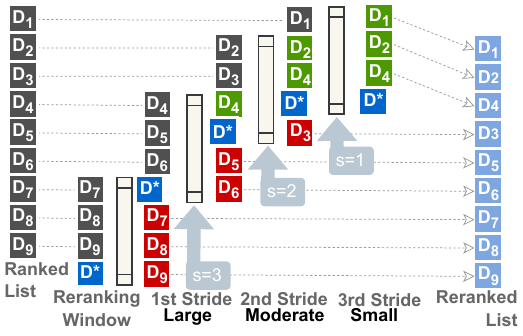}
    \caption{Workflow of comparison graph of a standard sliding window based algorithm with the inclusion of the generated pivot document ($D^*$) via the Gen-Sliding approach. It adapts the strides based on pivot-relative relevance density, taking longer 
    strides in sparse tail regions and shorter strides near the top.}
    \Description{Workflow diagram showing Gen-Sliding.}
    \label{fig:Gen-Sliding}
\end{figure}

\subsection{Generative Sliding Window}
\label{sec:gen-sliding-method}

The standard sliding-window strategy in Equation~\ref{eq:swindow} propagates
a fixed number of documents, controlled by the stride $s$, across consecutive
windows. We propose \textbf{Gen-Sliding}, a pivot-guided variant in which this
stride is determined dynamically for each window using the generated pivot
document $D^*$. The main idea is to apply the Gen-PART criterion with the listwise reranker locally
within each window: after inserting $D^*$ into the current window according to the ordering induced by the reranker $\phi$, only the documents
ranked above the pivot are propagated to the next window, as illustrated in Figure~\ref{fig:Gen-Sliding}.

More formally, consider a window $D_{[i:i+n]}$ of size $n$. We insert the
generated pivot document $D^*$ into this window. The ordering induced by the reranker $\phi$ yields the following pivot-induced ranking:
\begin{equation}
D^+_{[i:i+n]} =
\{D_j \in D_{[i:i+n]} : \phi(Q,D_j) \geq \phi(Q,D^*)\}.
\label{eq:gensliding-window-partition}
\end{equation}
That is, $D^+_{[i:i+n]}$ contains the documents in the current window that are
ranked above the pivot by the reranker. The number of documents
propagated from this window is then determined dynamically as
\begin{equation}
s_j = |D^+_{[i:i+n]}|,
\label{eq:gensliding-stride}
\end{equation}
rather than being fixed a priori by a global stride parameter. We restrict the maximum stride to $s_m$, where $s_m<n$, and define the final stride as $s_i=\min(s_m,s_j)$, thereby ensuring a bounded stride even when $s_j=n$.

Using notation analogous to the standard sliding-window computation graph in Equation~\ref{eq:swindow}, the Gen-Sliding workflow can be written as
\begin{equation}
\phi_{s_1}\!\left(D^+_{[m-n:m]}\right)
\rightarrow
\phi_{s_2}\!\left(D^+_{[m-n-s_1:m-s_1]}\right)
\cdots
\rightarrow
\phi_{s_T}\!\left(D^+_{[1:n]}\right),
\label{eq:gensliding}
\end{equation}
where $s_t$ denotes the number of documents propagated from the $t$-th
window and is determined by applying the pivot-based Gen-PART criterion with the listwise reranker within that window.

Thus, the Gen-Sliding comparisons (Equation \ref{eq:gensliding}) replace the fixed-stride propagation of conventional
sliding-window reranking (Equation \ref{eq:swindow}) with pivot-induced variable propagation. Intuitively, when the
pivot appears high in a window, only a small number of documents are likely to be retained, potentially leading to a larger effective step through the ranked list. Conversely, when the pivot appears lower in the window, more documents are likely to be propagated, potentially enabling more detailed reranking in regions where the pivot indicates higher local relevance density.

\begin{figure}[t]
    \centering
    \includegraphics[width=0.9\columnwidth]{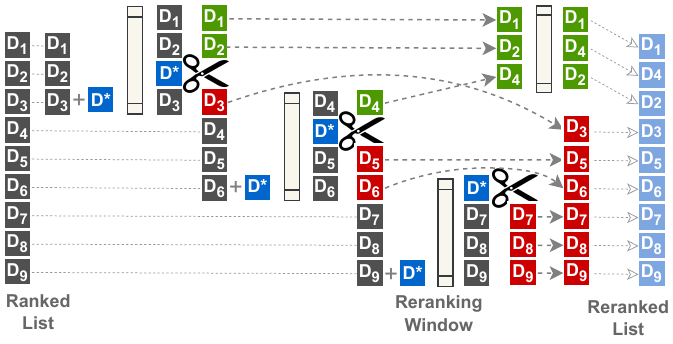}
    \caption{Gen-TDPart replaces the restricted pivot of the TDPart comparison algorithm \cite{parry2024topdownpartitioningefficientlistwise} (the top-scored document for the top sublist) with the generated relevant document $D^*$. This improves partition stability and ranking consistency.}
    \Description{Workflow diagram showing Gen-TDPart.}
    \label{fig:Gen-TDPart}
\end{figure}

\subsection{Generative Top-Down Partitioning}
\label{sec:gen-tdpart-method}

Top-Down Partitioning (\textbf{TDPart})~\cite{parry2024topdownpartitioningefficientlistwise}
improves the efficiency of listwise reranking by recursively partitioning the
candidate list using an anchor document selected from the top window. While
this strategy reduces the number of listwise comparisons, its effectiveness
depends critically on the quality of the selected anchor. If this anchor is
non-relevant or only weakly related to the query, the resulting partitions may
propagate retrieval noise throughout the subsequent reranking workflow.

We propose \textbf{Gen-TDPart}, a pivot-guided variant of TDPart that uses
$D^*$ as the partitioning anchor instead of selecting one from the initially
retrieved list. This makes the resulting comparison graph less sensitive to
retrieval noise, since partitioning is guided by a query-specific semantic
reference rather than by a potentially non-relevant retrieved document.
Following the pivot-based partitioning principle for pointwise rankers (Gen-PART), Gen-TDPart determines the rank position of the generated relevant document $D^*$ as induced by scores with the initial retriever model $\theta$. Subsequent top-down traversal is then guided by listwise comparisons involving the pivot
$D^*$.
Similar to the bottom-up approach of Section \ref{sec:gen-sliding-method}, for any sublist $D_{[a:b]}$ considered traversed sublist, we insert
$D^*$ into it and retain the documents ranked above the pivot:
\begin{equation}
D^+_{[a:b]} =
\{D_j \in D_{[a:b]} : \phi(Q, D_j) \succ \phi(Q, D^*)\},
\label{eq:gentdpart-partition}
\end{equation}
where $\succ$ denotes that the listwise reranker places $D_j$ above $D^*$ in the ordering induced over $D_{[a:b]} \cup \{D^*\}$. Documents ranked below the pivot are not promoted to the higher relevance partition (see Figure \ref{fig:Gen-TDPart} for an illustration).

Thus, Gen-TDPart preserves the top-down partitioning structure of TDPart while replacing its rank-derived anchor with a semantically grounded pivot. The resulting comparison graph is still efficient, since each partition is processed locally, but the partitioning decisions are now conditioned on the relative ordering of documents with respect to $D^*$. This allows document promotion to be governed by a query-specific semantic reference rather than by an anchor document selected by initial retrieval scores.

\begin{figure}[t]
    \centering
    \includegraphics[width=.65\columnwidth]{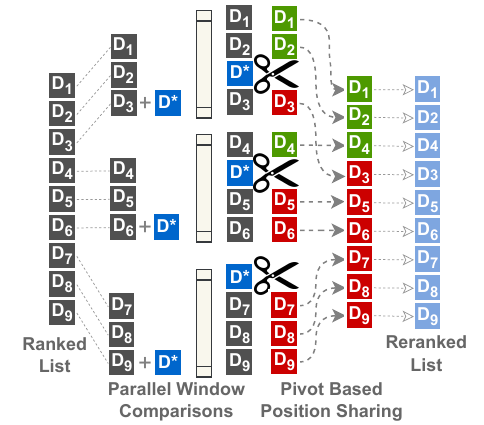}
    \caption{Our proposed listwise comparison graph approach SNOW,
    which processes non-overlapping windows in parallel, using $D^*$ (the pivot) to enable document movement across windows.
    }
    \Description{Workflow diagram showing SNOW.}
    \label{fig:SNOW_workflow}
\end{figure}

\subsection{Parallel Reranking using Pivot Document}
\label{sec:snow-method}

We propose a parallel comparison graph, \emph{\textbf{S}hared
\textbf{N}on-\textbf{O}verlapping \textbf{W}indow} (\textbf{SNOW}), which
processes disjoint windows in batches (see Figure~\ref{fig:SNOW_workflow}). Unlike standard sliding-window reranking, this removes overlap
between consecutive windows and eliminates sequential dependencies. However,
without overlap, documents cannot naturally propagate across windows. SNOW
addresses this limitation by inserting the shared pivot $D^*$ into every
window, providing a common semantic reference for all local listwise
comparisons. Thus, windows can be processed independently and in parallel (all batches at once when hardware resources permit), while their outputs remain comparable through their relative ordering with respect to $D^*$.
The ranked list is then partitioned into $m/n$ non-overlapping windows, each of
size $n$ -- $D_{[1:n]},\;
D_{[n+1:2n]},\;
\ldots,\;
D_{[m-n+1:m]}.
$
Applying the listwise reranker to each $\widehat{D}_{[a:b]}$ induces a
pivot-based partition, i.e.,
\begin{equation}
D^+_{[a:b]} =
\{D_j \in D_{[a:b]} : D_j \succ_{\phi}^{[a:b]} D^*\},\,\,
D^-_{[a:b]} = D_{[a:b]} \setminus D^+_{[a:b]} .
\label{eq:snow-window-partition}
\end{equation}
where
$\widehat{D}_{[a:b]} = D_{[a:b]} \cup \{D^*\}$ represents the pivot-augmented set,
and $\succ_{\phi}^{[a:b]}$ denotes the ordering produced by the ranker $\phi$ on $\widehat{D}_{[a:b]}$.
Documents in $D^+_{[a:b]}$ are promoted
relative to the pivot, whereas those in $D^-_{[a:b]}$ are demoted but not
discarded.
The SNOW comparison graph is then given by the following set of parallel
listwise inference calls:
\begin{equation}
\phi(Q,\widehat{D}_{[1:n]})
\,\parallel\,
\phi(Q,\widehat{D}_{[n+1:2n]})
\,\parallel\,
\cdots
\,\parallel\,
\phi(Q,\widehat{D}_{[m-n+1:m]}).
\label{eq:snow-graph}
\end{equation}

After all windows have been processed, the promoted and demoted portions are
concatenated separately across windows:
\begin{equation}
D^+ =
\bigoplus_{i=1}^{m/n}
D^+_{[(i-1)n+1:in]},\,\,
D^- =
\bigoplus_{i=1}^{m/n}
D^-_{[(i-1)n+1:in]},
\label{eq:snow-merge}
\end{equation}
where $\bigoplus$ denotes ordered list concatenation over the non-overlapping
windows. The final ordering is then obtained by placing the promoted list
before the demoted list:
\begin{equation}
L(Q; \theta, \phi) = D^+ \oplus D^-.
\label{eq:snow-final}
\end{equation}

\section{Experiment Setup}

\para{Research Questions}

We start with investigating whether modeling query-document relevance via the LLM-generated pivot (Section \ref{sec:sec3}) yields improvements in pointwise ranking (Section \ref{sec:gen-part})over standard RLT approaches that use supervised training on retrieval scores and collection statistics \cite{choppycut,Lien-Dynamic-RLT-2019}. Stated explicitly,
\uls
\li[] \textbf{RQ1:} Does modeling query-relevance via an LLM-generated document improve pointwise reranking performance?
\ule
Next, we investigate if the generated pivot document can be utilised effectively in existing listwise setups. More specifically,
\uls
\li[] \textbf{RQ2}: Does using LLM-generated documents as pivots improve a sliding window (Section \ref{sec:gen-sliding-method}) or a top-down partitioning reranking strategy (Section \ref{sec:gen-tdpart-method})?
\ule
Since the LLM-generated pivot document allows provision to make the sliding windows for a listwise setup non-overlapping and hence executable in parallel, the next research question is:
\uls
\li[] \textbf{RQ3}: Does the proposed listwise reranking method (i.e., SNOW of Section \ref{sec:snow-method}) lead to further improvements in trade-off between effectiveness and efficiency as compared to the adapted versions (Gen-Sliding and Gen-TDPart)?
\ule

\para{Datasets}

\begin{table}[t]
\centering
\caption{Dataset statistics for our experiments:
$\overline{|Q|}$ and $\overline{|D|}$ denote the average number
of query and document terms, respectively.}
\label{tab:dataset_stats}
\begin{adjustbox}{width=0.75\columnwidth}
\begin{tabular}{@{}lllrrr@{}}
\toprule
\textbf{Collection} & \textbf{\#Docs} & \textbf{Topics}
  & \textbf{\#Queries} & \textbf{$\overline{|Q|}$}
  & \textbf{$\overline{|D|}$} \\
\midrule
\textsc{MS}
  & \multirow{2}{*}{8.8M}
  & DL-19
  & 43
  & 5.40
  & \multirow{2}{*}{56.11} \\
\textsc{Marco}
  &  & DL-20 & 54 & 6.04 & \\
\midrule
\multirow{4}{*}{\textsc{BEIR}}
  & 171K  & COVID
  & 50    & 10.60
  & 160.77 \\
  & 5.2K  & SciFact & 300 & 12.37 & 213.63 \\
  & 383K  & Touché
  & 49    & 6.55
  & 292.37 \\
  & 4.6M  & DBPedia & 400 & 5.39 & 49.68 \\
\bottomrule
\end{tabular}
\end{adjustbox}
\end{table}

We evaluate our approach on the MS MARCO passage collection~\cite{bajaj2016ms} comprising over 8.8 million documents collated from the Bing search engine and then segmented into relatively short passages. For IR evaluation, we use the TREC deep learning track topics of 2019 and 2020, respectively denoted as DL-19~\cite{craswell2020overview} and DL-20~\cite{craswell2021overview}. We also conduct out-of-domain evaluation on a total of 4 datasets from the BEIR benchmark \cite{thakur2021beir} (summarised in Table \ref{tab:dataset_stats}).

\para{Evaluation Measures}
To balance precision towards higher ranks coupled with recall at deeper levels, we report the mean average precision (MAP) at cut-off 100 (\textbf{MAP@100}). The binary relevance threshold was set to 2 for DL-19 and DL-2020 topics as per the standard practice. To measure precision at higher ranks, we use \textbf{nDCG@10}, which is a standard evaluation measure modeling the behavior of a typical web search user \cite{ndcg} used for TREC DL and BEIR \cite{thakur2021beir}.

For pointwise rankers, efficiency is determined by the number of documents
passed to the reranker after applying RLT, i.e., the size of
$L(D^+,Q;\theta)$ in Equation~\ref{eq:dplus-and-dminus}. We therefore measure
the average truncation rank, which corresponds to the average number of
reranker inference calls per query. For each RLT method, we report the
efficiency gain as the ratio between the average number of inferences required
for full reranking and the average number required by the corresponding
truncation method.
For listwise reranking, efficiency is measured in terms of runtime. In
particular, we report the proportional runtime gain achieved by each
approximate sorting strategy relative to the standard sliding-window baseline,
which is the most computationally expensive approach considered in our
experiments. We refer to both quantities as the \textbf{efficiency gain ratio}
(\textbf{EGR}), where larger values indicate greater efficiency improvements. The cost of generating the pseudo-relevant document is included in the reported EGR for all of our proposed methods.

\para{Initial Retriever}
All our experiments, both in the pointwise and the listwise setups, require an initial retriever.
For evaluating our proposed pointwise reranking method (Gen-PART), we use only \textbf{BM25} as the first stage, as it is the most commonly used first-stage retriever used in existing RLT studies \cite{wu2021learningtruncaterankedlists}.
For LLM-based list-wise reranking setups, we use both \textbf{BM25} and \textbf{SPLADE} as the first-stage retriever. As the computational workflow of our generative listwise reranking methods depend on the rank positions assigned to the pivot document by the first-stage retriever across each iterative batch of documents, evaluating our approach on multiple retrievers investigates whether performance is consistent under different levels of initial retrieval quality.

\para{Pseudo-Relevant Document Generation}
Recall from Equation \ref{eq:pivotgen} that the summary document subsequently used as a pivot during reranking depends on an LLM $\psi$.
Our main results are reported with document pivots generated with a
moderate-sized $8$B parameters model \texttt{Llama-3.1-8B-Instruct}\footnote{\url{https://huggingface.co/meta-llama/Llama-3.1-8B-Instruct}}.
To investigate the sensitivity of our proposed approach on the choice of LLMs, we conduct additional experiments with two more generative models: a small model with $3$B parameters \texttt{Llama-3.2-3B-Instruct}\footnote{\url{https://huggingface.co/meta-llama/Llama-3.2-3B-Instruct}}, and a much larger model $\approx 200$B parameters \texttt{GPT-4o}\footnote{\url{https://platform.openai.com/docs/models/gpt-4o}}.

\para{Pointwise Reranking Setup}
As a common evaluation framework to compare Gen-PART with baseline RLT approaches, we experiment with two different multi-stage retriever configurations enlisted as follows.
\uls
\li
\textbf{BM25>>MT5:}
This is a 2-stage retrieve-rerank pipeline, where we employ an RLT method (Gen-PART or baselines) on top-100 BM25 retrieved documents, and then rerank the truncated list with mono-T5. In terms of our notations from Equation \ref{eq:rerankedlist}, this means $\phi(L(D^+,Q))$ corresponds to reranking the top-part of the list with mono-T5, whereas $L(D^-,Q)$ corresponds to the remainder of the BM25 retrieved list appended to the mono-T5 reranked one.

\li
\textbf{BM25>>MT5>>DT5:}
This is a three-stage retrieve-rerank pipeline, where we apply RLT in two stages. The first stage is identical to that of the previously listed setting, whereas the second stage involves applying duo-T5 \cite{pradeep2021expandomonoduodesignpatterntext} (a pairwise inference model) on the output of the top-50 documents of the first stage, i.e., as per our notation, in the second stage we rerank $m=50$ documents for a query $Q$ as ranked by $\theta \equiv \text{mono-T5}(\text{BM25}(Q))$.
\ule

\para{RLT Baseline approaches}
We compare our RLT approach against a number of standard supervised RLT baseline methods.
As per the experiment setting of \cite{meng2024ranked}, we use the TREC DL'19 topics to train these baseline RLT approaches (it is worth noting that our proposed method Gen-PART does not require a training set).
\uls

\li \textbf{Fixed}-$n$: truncates all queries at a pre-configured depth $n$. The setting specifically corresponding to $n=100$ means that no truncation is applied and thus refers to the max-effectiveness min-efficiency condition. EGR (efficiency-gain ratios) are hence reported relative to this setting.
\li \textbf{Greedy-$n$~\cite{Lien-Dynamic-RLT-2019}}:
This baseline also uses a fixed cutoff rank for all queries. However, rather
than setting this cutoff to an arbitrary pre-defined value, Greedy-$n$ selects
this cutoff value by maximizing an effectiveness--efficiency trade-off on a validation set (specifically, TREC DL'19 in our experiments) via grid search.
Following prior RLT study~\cite{meng2024ranked},
we define this trade-off using the weighted harmonic mean of effectiveness and efficiency. Specifically, as per~\cite{meng2024ranked} we quantify effectiveness as the difference of re-ranking
performance measured by nDCG@10 at a cut-off relative to the performance without reranking, whereas for efficiency, again as defined in \cite{meng2024ranked}, we used an exponentially decaying function of the cutoff rank, meaning smaller cutoffs yield higher efficiency.
As recommended by
\citeauthor{meng2024ranked}, we set the harmonic-mean parameter $\beta$ to 2.

\li \textbf{Supervised Dynamic RLT methods}:
This family of methods uses supervised training to model the functional dependence between the weighted F-score and rank cutoffs on a per-query basis. 
Among these while \textbf{BiCut}~\cite{Lien-Dynamic-RLT-2019} uses Bi-LSTMs, \textbf{Choppy}~\cite{choppycut} employs a transformer architecture to model the sequential dependence between the ranks and the effectiveness-efficiency ratio.
\textbf{AttnCut}~\cite{wu2021learningtruncaterankedlists} combines the two by utilizing a hybrid architecture of both Bi-LSTMs and transformers. \textbf{MtCut}~\cite{MtCut} extends AttnCut by employing multi-task learning to predict the relevance of each document in the ranked list while increasing
the margin between relevant and non-relevant documents.
Similar to the Greedy-$n$ approach, all these models use $\beta=2$.
\ule

\para{Listwise Reranking Setup}
For listwise reranking, we employ the following LLM-distilled approaches to rerank each window or sublist of documents across different iterations.
\uls
\li \textbf{Rank-Zephyr}: A list-wise LLM-based reranker designed to jointly assess and reorder a set of candidate documents in a single inference. It captures global ranking context and has demonstrated strong effectiveness in modern retrieval benchmarks~\cite{pradeep2023rankzephyreffectiverobustzeroshot}.

\li \textbf{Rank-Vicuna}: A list-wise reranker built on the Vicuna instruction-tuned LLM, utilizing conversational reasoning to perform holistic ranking over document lists. It provides competitive performance while remaining adaptable across ranking setups~\cite{pradeep2023rankvicunazeroshotlistwisedocument}.

\li \textbf{Rank-GPT}: A list-wise reranking comparison graph that utilizes large instruction-following language models to generate relevance-aware rankings over candidate lists. In our experiments, we use \texttt{GPT-4o-mini} for its balance between efficiency and effectiveness~\cite{sun2023chatgpt,Sinhababu-2025-ModelingRank}.
\ule

Since our proposed methods are pivot-based adaptations of the \textbf{Sliding} window \cite{Qin-2024-PRP} and the \textbf{TDPart} \cite{parry2024topdownpartitioningefficientlistwise} approaches, we employ both as baseline approximate sorting strategies for reranking the top-100 documents retrieved with BM25 and SPLADE. As per standard settings reported in \cite{Qin-2024-PRP,Sinhababu-2024-FSPRP}, we use a window size ($n$ of Equation \ref{eq:swindow}) of $20$ and a stride size of $10$. An identical window size of 20 is also used for TDPart. For SNOW, we use a batch size of $5$ (each batch containing $n=20$ documents). All the experiments are conducted in the same setting with an NVIDIA A100 40GB GPU.

\section{Results} \label{sec:results}

\definecolor{baselinecol}{RGB}{211, 211, 211}

\begin{table*}[t]
\centering
\caption{RLT-based reranking results on DL (in-domain) and BEIR (out-domain) on top-100 BM25 results. We report MAP@100 and nDCG@10 as effectiveness measures. Efficiency gain ratios (EGRs) are measured by reporting the average number of inferences per-query relative to the ``No-RLT'' setting (reranking all documents obtained by the BM25 retriever). We consider $n=20$ for the Fixed-$n$ baseline. Best results for each type of ranking pipeline are bold-faced.
\small
}
\label{tab:rlt_main}
\begin{adjustbox}{width=0.8\textwidth}
\begin{tabular}{@{}c l l l l l l l l l l l l l l l l@{}} 
\toprule
& & \multicolumn{3}{c}{DL-19 (train)} & \multicolumn{3}{c}{DL-20} & \multicolumn{2}{c}{COVID} & \multicolumn{2}{c}{SciFact} & \multicolumn{2}{c}{Touché} & \multicolumn{2}{c}{DBPedia}\\
\cmidrule(lr){3-5}
\cmidrule(lr){6-8}
\cmidrule(lr){9-10}
\cmidrule(lr){11-12}
\cmidrule(lr){13-14}
\cmidrule(lr){15-16}
& Ranker 
& MAP & nDCG & EGR 
& MAP & nDCG & EGR 
& nDCG & EGR 
& nDCG & EGR 
& nDCG & EGR 
& nDCG & EGR\\
\midrule

\multicolumn{2}{c}{1st-stage (BM25)}& .232 & .479 & n/a 
& .275 & .494 & n/a 
& .576 & n/a 
& .672 & n/a 
& .252 & n/a 
& .275 & n/a \\

\midrule
\multicolumn{16}{c}{1-stage Reranking: BM25>>MT5}\\
\midrule

\rowcolor{lightgray} 
No-RLT & MT5
& \textbf{.358} & \textbf{.701} & 1.00 
& \textbf{.374} & \textbf{.660} & 1.00 
& \textbf{.631} & 1.00 
& \textbf{.720} & 1.00 
& \textbf{.319} & 1.00 
& \textbf{.407} & 1.00 \\

Heuristic& Fixed 
& .300 & .605 & 5.00 
& .325 & .600 & 5.00 
& .620 & 5.00 
& .710 & 5.00 
& .295 & 5.00 
& .355 & 5.00 \\

RLT& Greedy 
& \textbf{.323} & .655 & 1.49 
& \textbf{.371} & .652 & 2.33 
& .623 & 2.33 
& .711 & 2.33 
& .315 & 2.33 
& .402 & 2.33 \\

\cmidrule{2-16}

& BiCut 
& .311 & .642 & 1.28 
& .354 & .640 & 1.54 
& .612 & 1.54 
& .698 & 1.54 
& .309 & 1.54 
& .395 & 1.54 \\

Supervised & Choppy 
& .321 & \textbf{.665} & 1.43 
& .368 & \textbf{.656} & 1.82 
& .627 & 1.82 
& .715 & 1.82 
& \textbf{.317} & 1.82 
& \textbf{.404} & 1.82 \\

RLT & AttnCut 
& .318 & .660 & 2.13 
& .364 & .650 & 2.78 
& .621 & \textbf{2.78} 
& .709 & 2.78 
& .314 & \textbf{2.78} 
& .401 & 2.79 \\

& MtCut 
& .308 & .655 & 2.26 
& .358 & .645 & \textbf{2.84} 
& .617 & \textbf{2.78} 
& .704 & 2.79
& .312 & \textbf{2.78} 
& .398 & 2.79 \\

\cmidrule{2-16}

Ours 
& Gen-PART 
& .315 & .634 & \textbf{2.71} 
& .327 & .630 & 2.57 
& \textbf{.635} & 2.29 
& \textbf{.716} & \textbf{7.69} 
& .295 & 1.98 
& .358 & \textbf{2.88} \\
\midrule
\multicolumn{16}{c}{2-stage reranking: BM25>>MT5>>DT5}\\
\midrule

\rowcolor{lightgray}
No-RLT & MT5>>DT5
& \textbf{.360} & \textbf{.703} & 1.00 
& \textbf{.385} & .666 & 1.00 
& \textbf{.631} & 1.00 
& \textbf{.579} & 1.00 
& \textbf{.272} & 1.00 
& \textbf{.404} & 1.00 \\

Heuristic& Fixed 
& .305 & .615 & 2.50 
& .330 & .605 & 2.50 
& .622 & 2.50 
& .565 & 2.50 
& .250 & 2.50 
& .358 & 2.50 \\

RLT& Greedy 
& .348 & .692 & 1.41 
& .372 & .655 & 1.90 
& .620 & 1.90 
& .568 & 1.82 
& .265 & 1.91 
& .395 & 1.91 \\

\cmidrule{2-16}

& BiCut 
& .312 & .668 & 1.24 
& .368 & .648 & 1.37 
& .618 & 1.44 
& .562 & 1.39 
& .262 & 1.45 
& .392 & 1.45 \\

Supervised& Choppy 
& .335 & .695 & 1.34 
& \textbf{.378} & \textbf{.658} & 1.60
& \textbf{.625} & 1.61 
& .572 & 1.57 
& .268 & 1.62 
& .399 & 1.63 \\

RLT & AttnCut 
& .332 & .690 & 1.71 
& .375 & .652 & 2.05 
& .622 & 2.11 
& .569 & 2.05 
& .266 & 2.12 
& .396 & 2.12 \\

& MtCut 
& .325 & .685 & 1.79 
& .370 & .650 & 2.13 
& .618 & 2.11 
& .565 & 2.05 
& .264 & 2.12 
& .394 & 2.12 \\

\cmidrule{2-16}

Ours 
& Gen-PART 
& \textbf{.360} & \textbf{.706} & \textbf{2.63} 
& .336 & .648 & \textbf{3.16} 
& .624 & \textbf{2.59} 
& \textbf{.669} & \textbf{7.28} 
& \textbf{.303} & \textbf{2.46} 
& \textbf{.416} & \textbf{3.52} \\
\bottomrule
\end{tabular}
\end{adjustbox}
\end{table*}

\label{ss:pointwise-results}

\para{Pointwise Reranking}
In relation to RQ1 (Gen-PART for RLT in pointwise ranking), we observe from Table~\ref{tab:rlt_main} that Gen-PART consistently achieves competitive or superior effectiveness (MAP@100 and nDCG@10) while requiring substantially reduced average number of reranker inferences to achieve this effectiveness, e.g., see that Gen-PART yields the highest EGR for DL-19 without a substantial decrease in MAP or nDCG relative to the best RLT baseline.

The improvements are particularly more pronounced on the 3-stage ranking pipeline where RLT is applied for the second time to further truncate a BM25>>mono-T5 result list with duo-T5 as the third-stage ranker, e.g., see that Gen-PART results in a 3.16$\times$ gain in efficiency relative to the full reranking setup with only a marginal drop in nDCG@10 relative to the full reranking pipeline (.648 vs. .666).
Similar trends, and in fact, gains of even higher magnitudes, are observed for out-of-domain evaluation of Gen-PART on the BEIR datasets, e.g., see how Gen-PART successfully results in an EGR 7.69 for SciFact. The likely reason for this is that Gen-PART does not involve any supervised training unlike the baseline RLT approaches which may overfit on the training topics.

\label{ss:listwise-results}

\begin{table}
\centering
\caption{Listwise LLM based reranking performance comparison on DL-19 and DL-20 datasets using the generated pivot for partitioning.
Each listwise ranker is a model distilled from their corresponding foundation model.
The best and the second-best scores for each metric and method are bold and \underline{underlined}. The symbol $*$ indicates the statistical significance of a method (paired t-test with $p = 0.05$) compared with the Sliding Window baseline.}
\label{tab:performance_comparison_dl19_dl20}
\resizebox{\columnwidth}{!}{%
\begin{tabular}{llllllll} 
\toprule
\multirow{2}{*}{Ranker} & \multirow{2}{*}{Pipeline} & \multicolumn{3}{c}{DL-19} & \multicolumn{3}{c}{DL-20} \\ 
\cmidrule(l){3-8}
 &  & MAP & nDCG & EGR & MAP & nDCG & EGR \\ 
\midrule
\rowcolor{lightgray}
Retriever & BM25 & .232 & .479 & – & .275 & .494 & – \\
\midrule
\multirow{5}{*}{Zephyr} 
 & + Sliding    & \underline{.342} & \textbf{.712} & 1.00 & \underline{.392} & \textbf{.705} & 1.00 \\
 & + Gen-Sliding & \textbf{.347} & \underline{.698} & \underline{1.67} & \textbf{.411} & \underline{.692} & \underline{1.73} \\
\cmidrule(lr){3-8}
 & + TDPart     & .279$^*$ & .641$^*$ & 1.46 & .345$^*$ & .659$^*$ & 1.48 \\ 
 & + Gen-TDPart  & .318 & .665$^*$ & 1.46 & .381 & .671$^*$ & 1.48 \\ 
\cmidrule(lr){2-8}
 & + SNOW        & .338 & .673$^*$ & \textbf{2.73} & .385 & .690 & \textbf{2.77} \\
\midrule
\multirow{5}{*}{Vicuna} 
 & + Sliding    & .285 & .586 & 1.00 & .315 & .577 & 1.00 \\
 & + Gen-Sliding & \textbf{.319} & \textbf{.653}$^*$ & \underline{1.67} & \textbf{.376}$^*$ & \textbf{.644}$^*$ & \underline{1.73} \\
\cmidrule(lr){3-8} 
 & + TDPart     & .235$^*$ & .525$^*$ & 1.46 & .260$^*$ & .520$^*$ & 1.48 \\ 
 & + Gen-TDPart  & \underline{.305} & \underline{.615} & 1.46 & \underline{.330} & \underline{.605} & 1.48 \\ 
\cmidrule(lr){2-8}
 & + SNOW        & .295 & .595 & \textbf{2.73} & .320 & .590 & \textbf{2.77} \\ 
\midrule
\multirow{5}{*}{GPT} 
 & + Sliding    & \underline{.345} & \textbf{.715} & 1.00 & \underline{.385} & \textbf{.693} & 1.00 \\
 & + Gen-Sliding & \textbf{.346} & \underline{.708} & \underline{1.67} & \textbf{.405} & \underline{.677} & \underline{1.73} \\
\cmidrule(lr){3-8}
 & + TDPart     & .284$^*$ & .638$^*$ & 1.46 & .322$^*$ & .628$^*$ & 1.48 \\ 
 & + Gen-TDPart  & .320 & .660$^*$ & 1.46 & .360 & .655$^*$ & 1.48 \\ 
\cmidrule(lr){2-8}
 & + SNOW        & .310 & .645$^*$ & \textbf{2.73} & .350 & .640$^*$ & \textbf{2.77} \\
\midrule
\rowcolor{lightgray}
Retriever & SPLADE & .485 & .730 & – & .505 & .721 & – \\
\midrule
\multirow{5}{*}{Zephyr} 
 & + Sliding    & \textbf{.545} & \textbf{.770} & 1.00 & \textbf{.585} & \underline{.799} & 1.00 \\
 & + Gen-Sliding & .461$^*$ & .747 & 1.45 & .555 & .766 & \underline{1.78} \\
\cmidrule(lr){3-8}
 & + TDPart     & .505 & .716$^*$ & \underline{1.46} & .540$^*$ & .773 & 1.48 \\ 
 & + Gen-TDPart  & \underline{.525} & .752 & \underline{1.46} & \underline{.560} & .785 & 1.48 \\ 
\cmidrule(lr){2-8}
 & + SNOW        & .489$^*$ & \underline{.767} & \textbf{2.73} & .544$^*$ & \textbf{.810} & \textbf{2.77} \\
\midrule
\multirow{5}{*}{Vicuna} 
 & + Sliding    & .462 & .712 & 1.00 & .468 & \underline{.706} & 1.00 \\
 & + Gen-Sliding & .451 & \textbf{.743}$^*$ & 1.45 & .486 & \textbf{.708} & \underline{1.78} \\
\cmidrule(lr){3-8}
 & + TDPart     & .430 & .662$^*$ & \underline{1.46} & .436 & .658$^*$ & 1.48 \\ 
 & + Gen-TDPart  & \underline{.502}$^*$ & .722 & \underline{1.46} & \underline{.525}$^*$ & .700 & 1.48 \\  
\cmidrule(lr){2-8}
 & + SNOW        & \textbf{.505}$^*$ & \underline{.725} & \textbf{2.73} & \textbf{.535}$^*$ & .704 & \textbf{2.77} \\
\midrule
\multirow{5}{*}{GPT} 
 & + Sliding    & .532 & .762 & 1.00 & .538 & .752 & 1.00 \\
 & + Gen-Sliding & .490 & \underline{.793}$^*$ & 1.45 & .542 & .769 & \underline{1.78} \\
\cmidrule(lr){3-8}
 & + TDPart     & .495 & .710$^*$ & 1.46 & .500 & .715$^*$ & 1.48 \\ 
 & + Gen-TDPart  & \underline{.545} & .775 & 1.46 & \underline{.565} & \underline{.770}$^*$ & 1.48 \\
\cmidrule(lr){2-8}
 & + SNOW        & \textbf{.560}$^*$ & \textbf{.795}$^*$ & \textbf{2.73} & \textbf{.570}$^*$ & \textbf{.780}$^*$ & \textbf{2.77} \\
\bottomrule
\end{tabular}}
\end{table}

\begin{table}[htbp!]
\centering
\caption{Listwise LLM based reranking performance comparison using RankZephyr on various out-of-domain datasets using the generated pivot for partitioning. The best and the second-best scores for each metric and method are bold and \underline{underlined}. The symbol $*$ indicates the statistical significance of a method (paired t-test with $p = 0.05$) compared with the Sliding Window baseline.}
\label{tab:performance_comparison_ood}
\begin{adjustbox}{width=\columnwidth}
\small
\begin{tabular}{@{}ll r ll r llr@{}}
\toprule
Pipeline
& \multicolumn{2}{c}{COVID} 
& \multicolumn{2}{c}{SciFact} 
& \multicolumn{2}{c}{Touché} 
& \multicolumn{2}{c}{DBPedia}\\

\cmidrule(lr){2-3} \cmidrule(lr){4-5} \cmidrule(lr){6-7} \cmidrule(lr){8-9}

& nDCG & EGR & nDCG & EGR & nDCG & EGR & nDCG & EGR\\
\midrule

BM25 & .576 & -- & .672 & --  & .252 & -- & .275 & -- \\
+ Sliding & .579 & 1.00 & .657 & 1.00 & \underline{.255} & 1.00 & \textbf{.410} & 1.00 \\
+ Gen-Sliding & \underline{.673}$^*$ & \underline{1.76} & \textbf{.747}$^*$ & \underline{1.82}  & \textbf{.266} & \underline{1.63} & \underline{.402} & \underline{1.93} \\
\cmidrule(lr){2-9}
+ TDPart & .598 & 1.55 & .649 & 1.55 & .209$^*$ & 1.46 & \underline{.383}$^*$ & 1.65 \\
+ Gen-TDPart & .610$^*$ & 1.55 & .660 & 1.55 & .220$^*$ & 1.46 & .395 & 1.65 \\
\cmidrule(lr){1-9}
+ SNOW & \textbf{.690}$^*$ & \textbf{2.80} & \underline{.730}$^*$ & \textbf{2.85} & .250 & \textbf{2.65} & .350$^*$ & \textbf{2.95} \\

\midrule

SPLADE & .531 & -- & .690 & -- & \textbf{.278} & -- & .426 & -- \\
+ Sliding & .536 & 1.00 & .677 & 1.00 & .274 & 1.00 & \underline{.472} & 1.00 \\
+ Gen-Sliding & \textbf{.641}$^*$ & \underline{1.81}  & \textbf{.764}$^*$ & \underline{1.80} & \textbf{.301}$^*$ & \underline{1.66} & .461 & \underline{1.98} \\
\cmidrule(lr){2-9}
+ TDPart & .562 & 1.55 & .684 & 1.55 & .246$^*$ & 1.46 & .445$^*$ & 1.65 \\
+ Gen-TDPart & .580$^*$ & 1.55 & .695 & 1.55 & .255 & 1.46 & .460 & 1.65 \\
\cmidrule(lr){1-9}
+ SNOW & \underline{.630}$^*$ & \textbf{2.80}  & \underline{.715}$^*$ & \textbf{2.85} & \underline{.275} & \textbf{2.65} & \textbf{.480} & \textbf{2.95} \\

\bottomrule
\end{tabular}
\end{adjustbox}
\end{table}

\para{Listwise Reranking}
In relation to RQ2, which examines whether the proposed pivot-based
adaptations of the comparison workflows for Sliding Window~\cite{Qin-2024-PRP}
and TDPart~\cite{parry2024topdownpartitioningefficientlistwise} improve over
their original counterparts, Table~\ref{tab:performance_comparison_dl19_dl20}
shows that Gen-Sliding consistently
reduces execution time relative to Sliding. This is reflected in EGR values
greater than 1 across all settings, indicating that the generated pivot is
effective in dynamically adjusting the stride according to its relative
position within each reranking window.
The effectiveness gains of Gen-Sliding, however, depend on the listwise
reranker used to position the pivot within each window. For example,
Gen-Sliding achieves higher MAP@100 and nDCG@10 than the original Sliding
strategy when using the RankVicuna reranker. This suggests that, when the
reranker can reliably compare candidate documents against the pivot, the
pivot-guided workflow can improve efficiency while also preserving, and in
some cases improving, ranking effectiveness.
The Gen-TDPart adaptation always outperforms its parent algorithm in terms of retrieval effectiveness while maintaining (almost) identical execution times.

In relation to RQ3, which evaluates whether the proposed approximate sorting
algorithm SNOW (Section~\ref{sec:snow-method}) improves over existing
comparison-workflow strategies, Table~\ref{tab:performance_comparison_dl19_dl20}
shows that SNOW achieves the largest efficiency gains. This improvement is
attributable to its pivot-based comparison graph, which allows the sublists
processed by the reranker to be evaluated in parallel
(Equation~\ref{eq:snow-final}).
Importantly, this substantial runtime reduction is accompanied by improvements
in retrieval effectiveness relative to other reranking strategies. For
example, when using Rank-GPT as the listwise reranker, SNOW achieves the best
MAP@100 and nDCG@10 results on both DL-19 and DL-20 when reranking the top-100
documents initially retrieved by SPLADE. These results indicate that the
pivot-guided non-overlapping window structure can improve efficiency while
also preserving, and in some cases enhancing, ranking effectiveness.
We observe similar trends for out-of-domain evaluation as well (results presented only for Rank-Zephyr in Table \ref{tab:performance_comparison_ood}), where we observe that 
SNOW, achieves the best efficiency (a maximum gain of $EGR=2.95$ on DBPedia) and demonstrates consistent performance across test collections. Under BM25 as a first-stage retriever, SNOW performs the best on COVID (nDCG of $.690$) and second best on SciFact (nDCG of $.730$).

\label{ss:llm-sensitivity-results}

\definecolor{gensliding}{HTML}{009E73}
\definecolor{gentdpart}{HTML}{D55E00}
\definecolor{snow}{HTML}{0072B2}

\begin{figure}[t]
\centering

\begin{tikzpicture}

\begin{groupplot}[
    group style={
        group size=2 by 1,
        horizontal sep=0.5cm
    },
    width=0.55\columnwidth,
    height=0.5\columnwidth,
    grid=major,
    grid style={dashed, opacity=0.35},
    xmin=0.8,
    xmax=3.2,
    ymin=0.62,
    ymax=0.85,
    xtick={1,2,3},
    xticklabels={Llama-3B,Llama-8B,GPT-4O},
    xlabel={Pivot Generator Models},
    tick label style={font=\scriptsize},
    label style={font=\scriptsize},
    title style={font=\scriptsize},
    legend style={
        font=\tiny,
        draw=none,
        at={(1,-0.4)},
        anchor=north,
        legend columns=5
    },
    every axis plot/.append style={
        line width=1pt,
        mark size=2.5pt,
    }
]

\nextgroupplot[
    title={DL-19},
    ylabel={nDCG@10},
    ymin=0.6,
    ymax=0.85,
]

\addlegendimage{color=black, mark=*, solid}
\addlegendentry{BM25}

\addlegendimage{color=black, mark=star, dashed}
\addlegendentry{SPLADE}

\addlegendimage{color=gensliding, mark=-}
\addlegendentry{Gen-Sliding}

\addlegendimage{color=gentdpart, mark=-}
\addlegendentry{Gen-TDPart}

\addlegendimage{color=snow, mark=-}
\addlegendentry{SNOW}

\addplot[
    color=gensliding,
    mark=*,
]
coordinates {(1,0.678) (2,0.697) (3,0.677)};

\addplot[
    color=gentdpart,
    mark=*
]
coordinates {(1,0.673) (2,0.664) (3,0.667)};

\addplot[
    color=snow,
    mark=*,
]
coordinates {(1,0.652) (2,0.672) (3,0.645)};

\addplot[
    color=gensliding,
    dashed,
    mark=star,
]
coordinates {(1,0.755) (2,0.747) (3,0.763)};

\addplot[
    color=gentdpart,
    dashed,
    mark=star,
]
coordinates {(1,0.758) (2,0.752) (3,0.753)};

\addplot[
    color=snow,
    dashed,
    mark=star,
]
coordinates {(1,0.775) (2,0.767) (3,0.769)};

\nextgroupplot[
    title={DL-20},
    yticklabels={,,},
    ymin=0.6,
    ymax=0.85
]

\addplot[
    color=gensliding,
    mark=*,
]
coordinates {(1,0.658) (2,0.691) (3,0.678)};

\addplot[
    color=gentdpart,
    mark=*,
]
coordinates {(1,0.672) (2,0.671) (3,0.659)};

\addplot[
    color=snow,
    mark=*,
]
coordinates {(1,0.670) (2,0.689) (3,0.657)};

\addplot[
    color=snow,
    dashed,
    mark=star,
]
coordinates {(1,0.809) (2,0.808) (3,0.797)};

\addplot[
    color=gensliding,
    dashed,
    mark=star,
]
coordinates {(1,0.767) (2,0.765) (3,0.758)};

\addplot[
    color=gentdpart,
    dashed,
    mark=star,
]
coordinates {(1,0.784) (2,0.783) (3,0.774)};

\end{groupplot}

\end{tikzpicture}

\caption{Our proposed listwise reranking is relatively insensitive to the LLMs used for pivot document generation (Equation \ref{eq:pivotgen}). Among the generation-based reranking approaches, Gen-TDPart exhibits the least variation in downstream retrieval effectiveness.}
\Description{Sensitivity of pseudo-relevant document generation across LLMs.}
\label{fig:llm_compare_pivot}

\end{figure}

\para{LLM Sensitivity}
The pivot document used for reranking with our proposed approaches in Tables \ref{tab:rlt_main}, \ref{tab:performance_comparison_dl19_dl20} and \ref{tab:performance_comparison_ood} were obtained with Llama-3.1, an 8B parameter model. 
Figure~\ref{fig:llm_compare_pivot} shows that ranking effectiveness exhibits minimal variation across pivots generated by other models, thus implying that the performance of the reranking workflow is relatively insensitive to the family or the size of an LLM.
Although we observe the maximum improvements with the larger model (GPT-4o), the cross-model differences remain negligible. Furthermore, the moderately sized model, Llama-3.1-8B-Instruct, provides a stable balance between the larger and the smaller models, making it a suitable choice for practical reranking setups.

\label{ss:ablation-of-rel-label}

\definecolor{dlzero}{HTML}{0072B2}
\definecolor{dltwenty}{HTML}{56B4E9}
\definecolor{covidcolor}{HTML}{009E73}
\definecolor{scifactcolor}{HTML}{CC79A7}
\definecolor{touchecolor}{HTML}{D55E00}
\definecolor{dbpediacolor}{HTML}{E69F00}

\begin{figure}[t]
\centering

\begin{tikzpicture}

\begin{groupplot}[
    group style={
        group size=2 by 1,
        horizontal sep=1cm
    },
    width=0.55\columnwidth,
    height=0.44\columnwidth,
    grid=major,
    grid style={opacity=0.3},
    xlabel={Relevance Threshold ($\tau$)},
    tick label style={font=\scriptsize},
    label style={font=\scriptsize},
    title style={font=\scriptsize},
    legend style={
        font=\tiny,
        draw=none,
        at={(0.5,-0.5)},
        anchor=west,
        legend columns=3
    },
    every axis plot/.append style={
        line width=0.9pt,
        mark size=2pt,
    }
]

\nextgroupplot[
    ylabel={nDCG@10},
    xtick={1,2,3},
    ymin=0.2,
    ymax=0.8,
]

\addplot[
    color=dlzero
]
coordinates {(1,0.681) (2,0.634) (3,0.579)};
\addlegendentry{DL-19}

\addplot[
    color=dltwenty
]
coordinates {(1,0.651) (2,0.630) (3,0.584)};
\addlegendentry{DL-20}

\addplot[
    color=covidcolor
]
coordinates {(1,0.633) (2,0.635) (3,0.612)};
\addlegendentry{COVID}

\addplot[
    color=scifactcolor
]
coordinates {(1,0.719) (2,0.716) (3,0.642)};
\addlegendentry{SciFact}

\addplot[
    color=touchecolor
]
coordinates {(1,0.308) (2,0.295) (3,0.267)};
\addlegendentry{Touché}

\addplot[
    color=dbpediacolor
]
coordinates {(1,0.388) (2,0.358) (3,0.312)};
\addlegendentry{DBPedia}

\nextgroupplot[
    ylabel={Average Cutoff},
    ylabel style={yshift=-5pt},
    xtick={3,2,1},
    ymin=0,
    ymax=100,
]

\addplot[
    color=dlzero
]
coordinates {(3,19.3) (2,44.2) (1,71.4)};

\addplot[
    color=dltwenty
]
coordinates {(3,17.8) (2,42.7) (1,69.3)};

\addplot[
    color=covidcolor
]
coordinates {(3,36.4) (2,71.8) (1,89.2)};

\addplot[
    color=scifactcolor
]
coordinates {(3,11.2) (2,23.6) (1,57.4)};

\addplot[
    color=touchecolor
]
coordinates {(3,33.1) (2,66.5) (1,87.3)};

\addplot[
    color=dbpediacolor
]
coordinates {(3,24.7) (2,51.3) (1,76.8)};

\end{groupplot}

\end{tikzpicture}

\caption{For Gen-PART, we observe that a pseudo-relevant document generated with $\tau=2$ yields a significantly lower average cutoff position (i.e., higher efficiency) than $\tau=1$, while maintaining performance. Furthermore, $\tau=3$, leads to smaller cutoffs (a small number of documents to rerank), resulting in a substantial drop in nDCG@10 effectiveness.}
\Description{Pseudo-relevant document generation by varying relevance signals.}
\label{fig:cutoff_performance_tau}

\end{figure}

\para{Sensitivity to Relevance Threshold for Pivot}
The performance and efficiency of our proposed methods, as presented in Tables \ref{tab:rlt_main}, \ref{tab:performance_comparison_dl19_dl20}, and \ref{tab:performance_comparison_ood}, depend on the generated pivot document and its relevance to the query. Our proposed pseudo-relevant document generation method (Section \ref{sec:sec3}) uses $\tau$ to specify the target relevance. We observe that $\tau=2$ acts as a boundary between relevant and non-relevant documents in the ranked list (Section \ref{ss:pointwise-results}), yielding a significantly lower cutoff position than $\tau=1$ while preserving nearly identical retrieval effectiveness as shown in Figure \ref{fig:cutoff_performance_tau}. In contrast, $\tau=1$ generates a non-relevant document, leading to higher cutoff positions and reduced efficiency gains, whereas $\tau=3$ corresponds to a highly relevant document, causing early cutoffs and a substantial drop in performance due to insufficient reranking depth. Based on this consistent trade-off, our $\tau=2$ setting balances effectiveness and efficiency in our experimental setup.

\section{Conclusions and Future Work}
Our empirical analysis confirms that LLM-generated pivot documents function as effective semantic thresholds for adaptive ranked list truncation (RLT) in both pointwise and pairwise reranking pipelines, achieving competitive MAP and nDCG@10 scores with 35--66\% inference reductions across in-domain and out-of-domain collections. Our proposed pivot-guided listwise reranking methods (SNOW, Gen-Sliding, and Gen-TDPart) consistently outperform baselines across test collections while showing stability across multiple LLMs (RankZephyr, RankVicuna, RankGPT), achieving significant efficiency gains while maintaining or improving performance. Pivot generation proves robust across different models, exhibiting minimal variance. These findings establish LLM-generated pivots as stable anchors that enhance multi-stage reranking efficiency and stability without model training, solely by leveraging LLMs' generative capabilities.

In the future, we plan to explore whether a collection-grounded generation of the pseudo-relevant documents may improve results. Another idea would be to work with multiple pivots, conditioning on the local document windows.

\bibliographystyle{ACM-Reference-Format.bst}
\bibliography{bibliography}


\begin{thebibliography}{44}


\ifx \showCODEN    \undefined \def \showCODEN     #1{\unskip}     \fi
\ifx \showISBNx    \undefined \def \showISBNx     #1{\unskip}     \fi
\ifx \showISBNxiii \undefined \def \showISBNxiii  #1{\unskip}     \fi
\ifx \showISSN     \undefined \def \showISSN      #1{\unskip}     \fi
\ifx \showLCCN     \undefined \def \showLCCN      #1{\unskip}     \fi
\ifx \shownote     \undefined \def \shownote      #1{#1}          \fi
\ifx \showarticletitle \undefined \def \showarticletitle #1{#1}   \fi
\ifx \showURL      \undefined \def \showURL       {\relax}        \fi
\providecommand\bibfield[2]{#2}
\providecommand\bibinfo[2]{#2}
\providecommand\natexlab[1]{#1}
\providecommand\showeprint[2][]{arXiv:#2}

\bibitem[Arabzadeh et~al\mbox{.}(2021)]%
        {arabzadeh2021bert}
\bibfield{author}{\bibinfo{person}{Negar Arabzadeh}, \bibinfo{person}{Maryam
  Khodabakhsh}, {and} \bibinfo{person}{Ebrahim Bagheri}.}
  \bibinfo{year}{2021}\natexlab{}.
\newblock \showarticletitle{BERT-QPP: Contextualized Pre-trained transformers
  for Query Performance Prediction}. In \bibinfo{booktitle}{\emph{Proceedings
  of the 30th ACM International Conference on Information \& Knowledge
  Management}} (Virtual Event, Queensland, Australia)
  \emph{(\bibinfo{series}{CIKM '21})}. \bibinfo{publisher}{Association for
  Computing Machinery}, \bibinfo{address}{New York, NY, USA},
  \bibinfo{pages}{2857–2861}.
\newblock
\showISBNx{9781450384469}
\href{https://doi.org/10.1145/3459637.3482063}{doi:\nolinkurl{10.1145/3459637.3482063}}


\bibitem[Arampatzis et~al\mbox{.}(2009)]%
        {where-to-stop}
\bibfield{author}{\bibinfo{person}{Avi Arampatzis}, \bibinfo{person}{Jaap
  Kamps}, {and} \bibinfo{person}{Stephen Robertson}.}
  \bibinfo{year}{2009}\natexlab{}.
\newblock \showarticletitle{Where to stop reading a ranked list? threshold
  optimization using truncated score distributions}. In
  \bibinfo{booktitle}{\emph{Proceedings of the 32nd International ACM SIGIR
  Conference on Research and Development in Information Retrieval}} (Boston,
  MA, USA) \emph{(\bibinfo{series}{SIGIR '09})}.
  \bibinfo{publisher}{Association for Computing Machinery},
  \bibinfo{address}{New York, NY, USA}, \bibinfo{pages}{524–531}.
\newblock
\showISBNx{9781605584836}
\href{https://doi.org/10.1145/1571941.1572031}{doi:\nolinkurl{10.1145/1571941.1572031}}


\bibitem[Bahri et~al\mbox{.}(2020)]%
        {choppycut}
\bibfield{author}{\bibinfo{person}{Dara Bahri}, \bibinfo{person}{Yi Tay},
  \bibinfo{person}{Che Zheng}, \bibinfo{person}{Donald Metzler}, {and}
  \bibinfo{person}{Andrew Tomkins}.} \bibinfo{year}{2020}\natexlab{}.
\newblock \showarticletitle{Choppy: Cut Transformer for Ranked List
  Truncation}. In \bibinfo{booktitle}{\emph{Proceedings of the 43rd
  International ACM SIGIR Conference on Research and Development in Information
  Retrieval}} (Virtual Event, China) \emph{(\bibinfo{series}{SIGIR '20})}.
  \bibinfo{publisher}{Association for Computing Machinery},
  \bibinfo{address}{New York, NY, USA}, \bibinfo{pages}{1513–1516}.
\newblock
\showISBNx{9781450380164}
\href{https://doi.org/10.1145/3397271.3401188}{doi:\nolinkurl{10.1145/3397271.3401188}}


\bibitem[Balog et~al\mbox{.}(2025)]%
        {Balog-LLMEval-2025}
\bibfield{author}{\bibinfo{person}{Krisztian Balog}, \bibinfo{person}{Don
  Metzler}, {and} \bibinfo{person}{Zhen Qin}.} \bibinfo{year}{2025}\natexlab{}.
\newblock \showarticletitle{Rankers, Judges, and Assistants: Towards
  Understanding the Interplay of LLMs in Information Retrieval Evaluation}. In
  \bibinfo{booktitle}{\emph{Proceedings of the 48th International ACM SIGIR
  Conference on Research and Development in Information Retrieval}} (Padua,
  Italy) \emph{(\bibinfo{series}{SIGIR '25})}. \bibinfo{publisher}{Association
  for Computing Machinery}, \bibinfo{address}{New York, NY, USA},
  \bibinfo{pages}{3865–3875}.
\newblock
\showISBNx{9798400715921}
\href{https://doi.org/10.1145/3726302.3730348}{doi:\nolinkurl{10.1145/3726302.3730348}}


\bibitem[Chen et~al\mbox{.}(2025)]%
        {chen-enhanceFIRST-2025}
\bibfield{author}{\bibinfo{person}{Zijian Chen}, \bibinfo{person}{Ronak
  Pradeep}, {and} \bibinfo{person}{Jimmy Lin}.}
  \bibinfo{year}{2025}\natexlab{}.
\newblock \showarticletitle{Accelerating Listwise Reranking: Reproducing and
  Enhancing FIRST}. In \bibinfo{booktitle}{\emph{Proceedings of the 48th
  International ACM SIGIR Conference on Research and Development in Information
  Retrieval}} (Padua, Italy) \emph{(\bibinfo{series}{SIGIR '25})}.
  \bibinfo{publisher}{Association for Computing Machinery},
  \bibinfo{address}{New York, NY, USA}, \bibinfo{pages}{3165–3172}.
\newblock
\showISBNx{9798400715921}
\href{https://doi.org/10.1145/3726302.3730287}{doi:\nolinkurl{10.1145/3726302.3730287}}


\bibitem[Craswell et~al\mbox{.}(2021)]%
        {craswell2021overview}
\bibfield{author}{\bibinfo{person}{Nick Craswell}, \bibinfo{person}{Bhaskar
  Mitra}, \bibinfo{person}{Emine Yilmaz}, {and} \bibinfo{person}{Daniel
  Campos}.} \bibinfo{year}{2021}\natexlab{}.
\newblock \bibinfo{title}{Overview of the TREC 2020 deep learning track}.
\newblock
\showeprint[arxiv]{2102.07662}~[cs.IR]


\bibitem[Craswell et~al\mbox{.}(2020)]%
        {craswell2020overview}
\bibfield{author}{\bibinfo{person}{Nick Craswell}, \bibinfo{person}{Bhaskar
  Mitra}, \bibinfo{person}{Emine Yilmaz}, \bibinfo{person}{Daniel Campos},
  {and} \bibinfo{person}{Ellen~M. Voorhees}.} \bibinfo{year}{2020}\natexlab{}.
\newblock \bibinfo{title}{Overview of the TREC 2019 deep learning track}.
\newblock
\showeprint[arxiv]{2003.07820}~[cs.IR]
\urldef\tempurl%
\url{https://arxiv.org/abs/2003.07820}
\showURL{%
\tempurl}


\bibitem[Datta et~al\mbox{.}(2022)]%
        {datta2022pointwise}
\bibfield{author}{\bibinfo{person}{Suchana Datta}, \bibinfo{person}{Sean
  MacAvaney}, \bibinfo{person}{Debasis Ganguly}, {and} \bibinfo{person}{Derek
  Greene}.} \bibinfo{year}{2022}\natexlab{}.
\newblock \showarticletitle{A 'Pointwise-Query, Listwise-Document' based Query
  Performance Prediction Approach}. In \bibinfo{booktitle}{\emph{Proceedings of
  the 45th International ACM SIGIR Conference on Research and Development in
  Information Retrieval}} (Madrid, Spain) \emph{(\bibinfo{series}{SIGIR '22})}.
  \bibinfo{publisher}{Association for Computing Machinery},
  \bibinfo{address}{New York, NY, USA}, \bibinfo{pages}{2148–2153}.
\newblock
\showISBNx{9781450387323}
\href{https://doi.org/10.1145/3477495.3531821}{doi:\nolinkurl{10.1145/3477495.3531821}}


\bibitem[Faggioli et~al\mbox{.}(2023)]%
        {Guglielmo-2023-Perspectives}
\bibfield{author}{\bibinfo{person}{Guglielmo Faggioli}, \bibinfo{person}{Laura
  Dietz}, \bibinfo{person}{Charles L.~A. Clarke}, \bibinfo{person}{Gianluca
  Demartini}, \bibinfo{person}{Matthias Hagen}, \bibinfo{person}{Claudia
  Hauff}, \bibinfo{person}{Noriko Kando}, \bibinfo{person}{Evangelos Kanoulas},
  \bibinfo{person}{Martin Potthast}, \bibinfo{person}{Benno Stein}, {and}
  \bibinfo{person}{Henning Wachsmuth}.} \bibinfo{year}{2023}\natexlab{}.
\newblock \showarticletitle{Perspectives on Large Language Models for Relevance
  Judgment}. In \bibinfo{booktitle}{\emph{Proceedings of the 2023 ACM SIGIR
  International Conference on Theory of Information Retrieval}} (Taipei,
  Taiwan) \emph{(\bibinfo{series}{ICTIR '23})}. \bibinfo{publisher}{Association
  for Computing Machinery}, \bibinfo{address}{New York, NY, USA},
  \bibinfo{pages}{39–50}.
\newblock
\showISBNx{9798400700736}
\href{https://doi.org/10.1145/3578337.3605136}{doi:\nolinkurl{10.1145/3578337.3605136}}


\bibitem[Gangi~Reddy et~al\mbox{.}(2024)]%
        {gangireddy_2024_first}
\bibfield{author}{\bibinfo{person}{Revanth Gangi~Reddy},
  \bibinfo{person}{JaeHyeok Doo}, \bibinfo{person}{Yifei Xu},
  \bibinfo{person}{Md~Arafat Sultan}, \bibinfo{person}{Deevya Swain},
  \bibinfo{person}{Avirup Sil}, {and} \bibinfo{person}{Heng Ji}.}
  \bibinfo{year}{2024}\natexlab{}.
\newblock \showarticletitle{{FIRST}: Faster Improved Listwise Reranking with
  Single Token Decoding}. In \bibinfo{booktitle}{\emph{Proceedings of the 2024
  Conference on Empirical Methods in Natural Language Processing}},
  \bibfield{editor}{\bibinfo{person}{Yaser Al-Onaizan}, \bibinfo{person}{Mohit
  Bansal}, {and} \bibinfo{person}{Yun-Nung Chen}} (Eds.).
  \bibinfo{publisher}{Association for Computational Linguistics},
  \bibinfo{address}{Miami, Florida, USA}, \bibinfo{pages}{8642--8652}.
\newblock
\href{https://doi.org/10.18653/v1/2024.emnlp-main.491}{doi:\nolinkurl{10.18653/v1/2024.emnlp-main.491}}


\bibitem[Gao et~al\mbox{.}(2023)]%
        {gao2022precise}
\bibfield{author}{\bibinfo{person}{Luyu Gao}, \bibinfo{person}{Xueguang Ma},
  \bibinfo{person}{Jimmy Lin}, {and} \bibinfo{person}{Jamie Callan}.}
  \bibinfo{year}{2023}\natexlab{}.
\newblock \showarticletitle{Precise Zero-Shot Dense Retrieval without Relevance
  Labels}. In \bibinfo{booktitle}{\emph{Proceedings of the 61st Annual Meeting
  of the Association for Computational Linguistics (Volume 1: Long Papers)}},
  \bibfield{editor}{\bibinfo{person}{Anna Rogers}, \bibinfo{person}{Jordan
  Boyd-Graber}, {and} \bibinfo{person}{Naoaki Okazaki}} (Eds.).
  \bibinfo{publisher}{Association for Computational Linguistics},
  \bibinfo{address}{Toronto, Canada}, \bibinfo{pages}{1762--1777}.
\newblock
\href{https://doi.org/10.18653/v1/2023.acl-long.99}{doi:\nolinkurl{10.18653/v1/2023.acl-long.99}}


\bibitem[Izacard et~al\mbox{.}(2022)]%
        {contriever}
\bibfield{author}{\bibinfo{person}{Gautier Izacard}, \bibinfo{person}{Mathilde
  Caron}, \bibinfo{person}{Lucas Hosseini}, \bibinfo{person}{Sebastian Riedel},
  \bibinfo{person}{Piotr Bojanowski}, \bibinfo{person}{Armand Joulin}, {and}
  \bibinfo{person}{Edouard Grave}.} \bibinfo{year}{2022}\natexlab{}.
\newblock \bibinfo{title}{Unsupervised Dense Information Retrieval with
  Contrastive Learning}.
\newblock
\showeprint[arxiv]{2112.09118}~[cs.IR]
\urldef\tempurl%
\url{https://arxiv.org/abs/2112.09118}
\showURL{%
\tempurl}


\bibitem[J\"{a}rvelin and Kek\"{a}l\"{a}inen(2002)]%
        {ndcg}
\bibfield{author}{\bibinfo{person}{Kalervo J\"{a}rvelin} {and}
  \bibinfo{person}{Jaana Kek\"{a}l\"{a}inen}.} \bibinfo{year}{2002}\natexlab{}.
\newblock \showarticletitle{Cumulated gain-based evaluation of IR techniques}.
\newblock \bibinfo{journal}{\emph{ACM Trans. Inf. Syst.}} \bibinfo{volume}{20},
  \bibinfo{number}{4} (\bibinfo{year}{2002}), \bibinfo{pages}{422–446}.
\newblock


\bibitem[Li et~al\mbox{.}(2025)]%
        {li2025leveragingreferencedocumentszeroshot}
\bibfield{author}{\bibinfo{person}{Jieran Li}, \bibinfo{person}{Xiuyuan Hu},
  \bibinfo{person}{Yang Zhao}, \bibinfo{person}{Shengyao Zhuang}, {and}
  \bibinfo{person}{Hao Zhang}.} \bibinfo{year}{2025}\natexlab{}.
\newblock \bibinfo{title}{Leveraging Reference Documents for Zero-Shot Ranking
  via Large Language Models}.
\newblock
\showeprint[arxiv]{2506.11452}~[cs.IR]
\urldef\tempurl%
\url{https://arxiv.org/abs/2506.11452}
\showURL{%
\tempurl}


\bibitem[Lien et~al\mbox{.}(2019)]%
        {Lien-Dynamic-RLT-2019}
\bibfield{author}{\bibinfo{person}{Yen-Chieh Lien}, \bibinfo{person}{Daniel
  Cohen}, {and} \bibinfo{person}{W.~Bruce Croft}.}
  \bibinfo{year}{2019}\natexlab{}.
\newblock \showarticletitle{An Assumption-Free Approach to the Dynamic
  Truncation of Ranked Lists}. In \bibinfo{booktitle}{\emph{Proceedings of the
  2019 ACM SIGIR International Conference on Theory of Information Retrieval}}
  (Santa Clara, CA, USA) \emph{(\bibinfo{series}{ICTIR '19})}.
  \bibinfo{publisher}{Association for Computing Machinery},
  \bibinfo{address}{New York, NY, USA}, \bibinfo{pages}{79–82}.
\newblock
\showISBNx{9781450368810}
\href{https://doi.org/10.1145/3341981.3344234}{doi:\nolinkurl{10.1145/3341981.3344234}}


\bibitem[Liu et~al\mbox{.}(2024)]%
        {liu2024slidingwindowsendexploring}
\bibfield{author}{\bibinfo{person}{Wenhan Liu}, \bibinfo{person}{Xinyu Ma},
  \bibinfo{person}{Yutao Zhu}, \bibinfo{person}{Ziliang Zhao},
  \bibinfo{person}{Shuaiqiang Wang}, \bibinfo{person}{Dawei Yin}, {and}
  \bibinfo{person}{Zhicheng Dou}.} \bibinfo{year}{2024}\natexlab{}.
\newblock \bibinfo{title}{Sliding Windows Are Not the End: Exploring Full
  Ranking with Long-Context Large Language Models}.
\newblock
\showeprint[arxiv]{2412.14574}~[cs.IR]
\urldef\tempurl%
\url{https://arxiv.org/abs/2412.14574}
\showURL{%
\tempurl}


\bibitem[Luo et~al\mbox{.}(2024)]%
        {luo-etal-2024-prp}
\bibfield{author}{\bibinfo{person}{Jian Luo}, \bibinfo{person}{Xuanang Chen},
  \bibinfo{person}{Ben He}, {and} \bibinfo{person}{Le Sun}.}
  \bibinfo{year}{2024}\natexlab{}.
\newblock \showarticletitle{{PRP}-Graph: Pairwise Ranking Prompting to {LLM}s
  with Graph Aggregation for Effective Text Re-ranking}. In
  \bibinfo{booktitle}{\emph{Proceedings of the 62nd Annual Meeting of the
  Association for Computational Linguistics (Volume 1: Long Papers)}},
  \bibfield{editor}{\bibinfo{person}{Lun-Wei Ku}, \bibinfo{person}{Andre
  Martins}, {and} \bibinfo{person}{Vivek Srikumar}} (Eds.).
  \bibinfo{publisher}{Association for Computational Linguistics},
  \bibinfo{address}{Bangkok, Thailand}, \bibinfo{pages}{5766--5776}.
\newblock
\href{https://doi.org/10.18653/v1/2024.acl-long.313}{doi:\nolinkurl{10.18653/v1/2024.acl-long.313}}


\bibitem[Ma et~al\mbox{.}(2023)]%
        {ma2023fine}
\bibfield{author}{\bibinfo{person}{Xueguang Ma}, \bibinfo{person}{Xinyu Zhang},
  \bibinfo{person}{Ronak Pradeep}, {and} \bibinfo{person}{Jimmy Lin}.}
  \bibinfo{year}{2023}\natexlab{}.
\newblock \showarticletitle{Fine-Tuning {LLaMA} for Multi-Stage Text
  Retrieval}. In \bibinfo{booktitle}{\emph{{SIGIR} '23: Proceedings of the 46th
  International ACM SIGIR Conference on Research and Development in Information
  Retrieval}}. \bibinfo{publisher}{{ACM}}, \bibinfo{address}{Taipei, Taiwan},
  \bibinfo{pages}{2659--2669}.
\newblock
\urldef\tempurl%
\url{https://arxiv.org/abs/2310.08319}
\showURL{%
\tempurl}


\bibitem[Ma et~al\mbox{.}(2022)]%
        {RetrievalInformation}
\bibfield{author}{\bibinfo{person}{Yixiao Ma}, \bibinfo{person}{Qingyao Ai},
  \bibinfo{person}{Yueyue Wu}, \bibinfo{person}{Yunqiu Shao},
  \bibinfo{person}{Yiqun Liu}, \bibinfo{person}{Min Zhang}, {and}
  \bibinfo{person}{Shaoping Ma}.} \bibinfo{year}{2022}\natexlab{}.
\newblock \showarticletitle{Incorporating Retrieval Information into the
  Truncation of Ranking Lists for Better Legal Search}. In
  \bibinfo{booktitle}{\emph{Proceedings of the 45th International ACM SIGIR
  Conference on Research and Development in Information Retrieval}} (Madrid,
  Spain) \emph{(\bibinfo{series}{SIGIR '22})}. \bibinfo{publisher}{Association
  for Computing Machinery}, \bibinfo{address}{New York, NY, USA},
  \bibinfo{pages}{438–448}.
\newblock
\showISBNx{9781450387323}
\href{https://doi.org/10.1145/3477495.3531998}{doi:\nolinkurl{10.1145/3477495.3531998}}


\bibitem[Meng et~al\mbox{.}(2024)]%
        {meng2024ranked}
\bibfield{author}{\bibinfo{person}{Chuan Meng}, \bibinfo{person}{Negar
  Arabzadeh}, \bibinfo{person}{Arian Askari}, \bibinfo{person}{Mohammad
  Aliannejadi}, {and} \bibinfo{person}{Maarten de Rijke}.}
  \bibinfo{year}{2024}\natexlab{}.
\newblock \showarticletitle{Ranked List Truncation for Large Language
  Model-based Re-Ranking}. In \bibinfo{booktitle}{\emph{Proceedings of the 47th
  International ACM SIGIR Conference on Research and Development in Information
  Retrieval}} (Washington DC, USA) \emph{(\bibinfo{series}{SIGIR '24})}.
  \bibinfo{publisher}{Association for Computing Machinery},
  \bibinfo{address}{New York, NY, USA}, \bibinfo{pages}{141–151}.
\newblock
\showISBNx{9798400704314}
\href{https://doi.org/10.1145/3626772.3657864}{doi:\nolinkurl{10.1145/3626772.3657864}}


\bibitem[Meng et~al\mbox{.}(2026)]%
        {meng-2026-rerankersrelevancejudges}
\bibfield{author}{\bibinfo{person}{Chuan Meng}, \bibinfo{person}{Jiqun Liu},
  \bibinfo{person}{Mohammad Aliannejadi}, \bibinfo{person}{Fengran Mo},
  \bibinfo{person}{Jeff Dalton}, {and} \bibinfo{person}{Maarten de Rijke}.}
  \bibinfo{year}{2026}\natexlab{}.
\newblock \bibinfo{title}{Re-Rankers as Relevance Judges}.
\newblock
\showeprint[arxiv]{2601.04455}~[cs.IR]
\urldef\tempurl%
\url{https://arxiv.org/abs/2601.04455}
\showURL{%
\tempurl}


\bibitem[Nguyen et~al\mbox{.}(2017)]%
        {bajaj2016ms}
\bibfield{author}{\bibinfo{person}{Tri Nguyen}, \bibinfo{person}{Mir
  Rosenberg}, \bibinfo{person}{Xia Song}, \bibinfo{person}{Jianfeng Gao},
  \bibinfo{person}{Saurabh Tiwary}, \bibinfo{person}{Rangan Majumder}, {and}
  \bibinfo{person}{Li Deng}.} \bibinfo{year}{2017}\natexlab{}.
\newblock \bibinfo{title}{{MS} {MARCO}: A Human-Generated {MA}chine Reading
  {CO}mprehension Dataset}.
\newblock
\urldef\tempurl%
\url{https://openreview.net/forum?id=Hk1iOLcle}
\showURL{%
\tempurl}


\bibitem[Nogueira and Cho(2019)]%
        {nogueira2019passage}
\bibfield{author}{\bibinfo{person}{Rodrigo Nogueira} {and}
  \bibinfo{person}{Kyunghyun Cho}.} \bibinfo{year}{2019}\natexlab{}.
\newblock \bibinfo{title}{Passage Re-ranking with {BERT}}.
\newblock \bibinfo{howpublished}{\url{https://arxiv.org/abs/1901.04085}}.
\newblock


\bibitem[Parry et~al\mbox{.}(2024)]%
        {parry2024topdownpartitioningefficientlistwise}
\bibfield{author}{\bibinfo{person}{Andrew Parry}, \bibinfo{person}{Sean
  MacAvaney}, {and} \bibinfo{person}{Debasis Ganguly}.}
  \bibinfo{year}{2024}\natexlab{}.
\newblock \bibinfo{title}{Top-Down Partitioning for Efficient List-Wise
  Ranking}.
\newblock
\showeprint[arxiv]{2405.14589}~[cs.IR]
\urldef\tempurl%
\url{https://arxiv.org/abs/2405.14589}
\showURL{%
\tempurl}


\bibitem[Pradeep et~al\mbox{.}(2021)]%
        {pradeep2021expandomonoduodesignpatterntext}
\bibfield{author}{\bibinfo{person}{Ronak Pradeep}, \bibinfo{person}{Rodrigo
  Nogueira}, {and} \bibinfo{person}{Jimmy Lin}.}
  \bibinfo{year}{2021}\natexlab{}.
\newblock \bibinfo{title}{The Expando-Mono-Duo Design Pattern for Text Ranking
  with Pretrained Sequence-to-Sequence Models}.
\newblock
\showeprint[arxiv]{2101.05667}~[cs.IR]
\urldef\tempurl%
\url{https://arxiv.org/abs/2101.05667}
\showURL{%
\tempurl}


\bibitem[Pradeep et~al\mbox{.}(2023a)]%
        {pradeep2023rankvicunazeroshotlistwisedocument}
\bibfield{author}{\bibinfo{person}{Ronak Pradeep}, \bibinfo{person}{Sahel
  Sharifymoghaddam}, {and} \bibinfo{person}{Jimmy Lin}.}
  \bibinfo{year}{2023}\natexlab{a}.
\newblock \bibinfo{title}{RankVicuna: Zero-Shot Listwise Document Reranking
  with Open-Source Large Language Models}.
\newblock
\showeprint[arxiv]{2309.15088}~[cs.IR]
\urldef\tempurl%
\url{https://arxiv.org/abs/2309.15088}
\showURL{%
\tempurl}


\bibitem[Pradeep et~al\mbox{.}(2023b)]%
        {pradeep2023rankzephyreffectiverobustzeroshot}
\bibfield{author}{\bibinfo{person}{Ronak Pradeep}, \bibinfo{person}{Sahel
  Sharifymoghaddam}, {and} \bibinfo{person}{Jimmy Lin}.}
  \bibinfo{year}{2023}\natexlab{b}.
\newblock \bibinfo{title}{RankZephyr: Effective and Robust Zero-Shot Listwise
  Reranking is a Breeze!}
\newblock
\showeprint[arxiv]{2312.02724}~[cs.IR]
\urldef\tempurl%
\url{https://arxiv.org/abs/2312.02724}
\showURL{%
\tempurl}


\bibitem[Qin et~al\mbox{.}(2024)]%
        {Qin-2024-PRP}
\bibfield{author}{\bibinfo{person}{Zhen Qin}, \bibinfo{person}{Rolf Jagerman},
  \bibinfo{person}{Kai Hui}, \bibinfo{person}{Honglei Zhuang},
  \bibinfo{person}{Junru Wu}, \bibinfo{person}{Le Yan},
  \bibinfo{person}{Jiaming Shen}, \bibinfo{person}{Tianqi Liu},
  \bibinfo{person}{Jialu Liu}, \bibinfo{person}{Donald Metzler},
  \bibinfo{person}{Xuanhui Wang}, {and} \bibinfo{person}{Michael Bendersky}.}
  \bibinfo{year}{2024}\natexlab{}.
\newblock \showarticletitle{Large Language Models are Effective Text Rankers
  with Pairwise Ranking Prompting}. In \bibinfo{booktitle}{\emph{Findings of
  the Association for Computational Linguistics: NAACL 2024}},
  \bibfield{editor}{\bibinfo{person}{Kevin Duh}, \bibinfo{person}{Helena
  Gomez}, {and} \bibinfo{person}{Steven Bethard}} (Eds.).
  \bibinfo{publisher}{Association for Computational Linguistics},
  \bibinfo{address}{Mexico City, Mexico}, \bibinfo{pages}{1504--1518}.
\newblock
\href{https://doi.org/10.18653/v1/2024.findings-naacl.97}{doi:\nolinkurl{10.18653/v1/2024.findings-naacl.97}}


\bibitem[Rahmani et~al\mbox{.}(2024)]%
        {Rahmani-2024-LLM4Eval}
\bibfield{author}{\bibinfo{person}{Hossein~A. Rahmani},
  \bibinfo{person}{Clemencia Siro}, \bibinfo{person}{Mohammad Aliannejadi},
  \bibinfo{person}{Nick Craswell}, \bibinfo{person}{Charles L.~A. Clarke},
  \bibinfo{person}{Guglielmo Faggioli}, \bibinfo{person}{Bhaskar Mitra},
  \bibinfo{person}{Paul Thomas}, {and} \bibinfo{person}{Emine Yilmaz}.}
  \bibinfo{year}{2024}\natexlab{}.
\newblock \showarticletitle{LLM4Eval: Large Language Model for Evaluation in
  IR}. In \bibinfo{booktitle}{\emph{Proceedings of the 47th International ACM
  SIGIR Conference on Research and Development in Information Retrieval}}
  (Washington DC, USA) \emph{(\bibinfo{series}{SIGIR '24})}.
  \bibinfo{publisher}{Association for Computing Machinery},
  \bibinfo{address}{New York, NY, USA}, \bibinfo{pages}{3040–3043}.
\newblock
\showISBNx{9798400704314}
\href{https://doi.org/10.1145/3626772.3657992}{doi:\nolinkurl{10.1145/3626772.3657992}}


\bibitem[Ren et~al\mbox{.}(2025)]%
        {Ruiyang-SelfCalibratedListwiseRerankingLLM-2025}
\bibfield{author}{\bibinfo{person}{Ruiyang Ren}, \bibinfo{person}{Yuhao Wang},
  \bibinfo{person}{Kun Zhou}, \bibinfo{person}{Wayne~Xin Zhao},
  \bibinfo{person}{Wenjie Wang}, \bibinfo{person}{Jing Liu},
  \bibinfo{person}{Ji-Rong Wen}, {and} \bibinfo{person}{Tat-Seng Chua}.}
  \bibinfo{year}{2025}\natexlab{}.
\newblock \showarticletitle{Self-Calibrated Listwise Reranking with Large
  Language Models}. In \bibinfo{booktitle}{\emph{Proceedings of the ACM on Web
  Conference 2025}} (Sydney NSW, Australia) \emph{(\bibinfo{series}{WWW '25})}.
  \bibinfo{publisher}{Association for Computing Machinery},
  \bibinfo{address}{New York, NY, USA}, \bibinfo{pages}{3692–3701}.
\newblock
\showISBNx{9798400712746}
\href{https://doi.org/10.1145/3696410.3714658}{doi:\nolinkurl{10.1145/3696410.3714658}}


\bibitem[Robertson et~al\mbox{.}(1995)]%
        {robertson1995okapi}
\bibfield{author}{\bibinfo{person}{Stephen~E Robertson}, \bibinfo{person}{Steve
  Walker}, \bibinfo{person}{Susan Jones}, \bibinfo{person}{Micheline~M
  Hancock-Beaulieu}, {and} \bibinfo{person}{Mike Gatford}.}
  \bibinfo{year}{1995}\natexlab{}.
\newblock \showarticletitle{Okapi at TREC-3}. In
  \bibinfo{booktitle}{\emph{Overview of the Third Text REtrieval Conference
  (TREC-3)}}. \bibinfo{publisher}{NIST Special Publication 500-225},
  \bibinfo{address}{Gaithersburg, MD}, \bibinfo{pages}{109--126}.
\newblock


\bibitem[Singh et~al\mbox{.}(2023)]%
        {singh2023unsupervised}
\bibfield{author}{\bibinfo{person}{Ashutosh Singh}, \bibinfo{person}{Debasis
  Ganguly}, \bibinfo{person}{Suchana Datta}, {and} \bibinfo{person}{Craig
  McDonald}.} \bibinfo{year}{2023}\natexlab{}.
\newblock \showarticletitle{Unsupervised Query Performance Prediction for
  Neural Models with Pairwise Rank Preferences}. In
  \bibinfo{booktitle}{\emph{Proceedings of the 46th International ACM SIGIR
  Conference on Research and Development in Information Retrieval}} (Taipei,
  Taiwan) \emph{(\bibinfo{series}{SIGIR '23})}. \bibinfo{publisher}{Association
  for Computing Machinery}, \bibinfo{address}{New York, NY, USA},
  \bibinfo{pages}{2486–2490}.
\newblock
\showISBNx{9781450394086}
\href{https://doi.org/10.1145/3539618.3592082}{doi:\nolinkurl{10.1145/3539618.3592082}}


\bibitem[Sinhababu et~al\mbox{.}(2025)]%
        {Sinhababu-2025-ModelingRank}
\bibfield{author}{\bibinfo{person}{Nilanjan Sinhababu}, \bibinfo{person}{Andrew
  Parry}, \bibinfo{person}{Debasis Ganguly}, {and} \bibinfo{person}{Pabitra
  Mitra}.} \bibinfo{year}{2025}\natexlab{}.
\newblock \bibinfo{title}{Modeling Ranking Properties with In-Context
  Learning}.
\newblock
\showeprint[arxiv]{2505.17736}~[cs.IR]
\urldef\tempurl%
\url{https://arxiv.org/abs/2505.17736}
\showURL{%
\tempurl}


\bibitem[Sinhababu et~al\mbox{.}(2024)]%
        {Sinhababu-2024-FSPRP}
\bibfield{author}{\bibinfo{person}{Nilanjan Sinhababu}, \bibinfo{person}{Andrew
  Parry}, \bibinfo{person}{Debasis Ganguly}, \bibinfo{person}{Debasis Samanta},
  {and} \bibinfo{person}{Pabitra Mitra}.} \bibinfo{year}{2024}\natexlab{}.
\newblock \showarticletitle{Few-shot Prompting for Pairwise Ranking: An
  Effective Non-Parametric Retrieval Model}. In
  \bibinfo{booktitle}{\emph{Findings of the Association for Computational
  Linguistics: EMNLP 2024}}, \bibfield{editor}{\bibinfo{person}{Yaser
  Al-Onaizan}, \bibinfo{person}{Mohit Bansal}, {and} \bibinfo{person}{Yun-Nung
  Chen}} (Eds.). \bibinfo{publisher}{Association for Computational
  Linguistics}, \bibinfo{address}{Miami, Florida, USA},
  \bibinfo{pages}{12363--12377}.
\newblock
\href{https://doi.org/10.18653/v1/2024.findings-emnlp.720}{doi:\nolinkurl{10.18653/v1/2024.findings-emnlp.720}}


\bibitem[Sun et~al\mbox{.}(2023)]%
        {sun2023chatgpt}
\bibfield{author}{\bibinfo{person}{Weiwei Sun}, \bibinfo{person}{Lingyong Yan},
  \bibinfo{person}{Xinyu Ma}, \bibinfo{person}{Shuaiqiang Wang},
  \bibinfo{person}{Pengjie Ren}, \bibinfo{person}{Zhumin Chen},
  \bibinfo{person}{Dawei Yin}, {and} \bibinfo{person}{Zhaochun Ren}.}
  \bibinfo{year}{2023}\natexlab{}.
\newblock \showarticletitle{Is {C}hat{GPT} Good at Search? Investigating Large
  Language Models as Re-Ranking Agents}. In
  \bibinfo{booktitle}{\emph{Proceedings of the 2023 Conference on Empirical
  Methods in Natural Language Processing}},
  \bibfield{editor}{\bibinfo{person}{Houda Bouamor}, \bibinfo{person}{Juan
  Pino}, {and} \bibinfo{person}{Kalika Bali}} (Eds.).
  \bibinfo{publisher}{Association for Computational Linguistics},
  \bibinfo{address}{Singapore}, \bibinfo{pages}{14918--14937}.
\newblock
\href{https://doi.org/10.18653/v1/2023.emnlp-main.923}{doi:\nolinkurl{10.18653/v1/2023.emnlp-main.923}}


\bibitem[Tang et~al\mbox{.}(2024)]%
        {tang_2024_found}
\bibfield{author}{\bibinfo{person}{Raphael Tang}, \bibinfo{person}{Crystina
  Zhang}, \bibinfo{person}{Xueguang Ma}, \bibinfo{person}{Jimmy Lin}, {and}
  \bibinfo{person}{Ferhan Ture}.} \bibinfo{year}{2024}\natexlab{}.
\newblock \showarticletitle{Found in the Middle: Permutation Self-Consistency
  Improves Listwise Ranking in Large Language Models}. In
  \bibinfo{booktitle}{\emph{Proceedings of the 2024 Conference of the North
  American Chapter of the Association for Computational Linguistics: Human
  Language Technologies (Volume 1: Long Papers)}},
  \bibfield{editor}{\bibinfo{person}{Kevin Duh}, \bibinfo{person}{Helena
  Gomez}, {and} \bibinfo{person}{Steven Bethard}} (Eds.).
  \bibinfo{publisher}{Association for Computational Linguistics},
  \bibinfo{address}{Mexico City, Mexico}, \bibinfo{pages}{2327--2340}.
\newblock
\href{https://doi.org/10.18653/v1/2024.naacl-long.129}{doi:\nolinkurl{10.18653/v1/2024.naacl-long.129}}


\bibitem[Thakur et~al\mbox{.}(2021)]%
        {thakur2021beir}
\bibfield{author}{\bibinfo{person}{Nandan Thakur}, \bibinfo{person}{Nils
  Reimers}, \bibinfo{person}{Andreas Rücklé}, \bibinfo{person}{Abhishek
  Srivastava}, {and} \bibinfo{person}{Iryna Gurevych}.}
  \bibinfo{year}{2021}\natexlab{}.
\newblock \bibinfo{title}{BEIR: A Heterogenous Benchmark for Zero-shot
  Evaluation of Information Retrieval Models}.
\newblock
\showeprint[arxiv]{2104.08663}~[cs.IR]
\urldef\tempurl%
\url{https://arxiv.org/abs/2104.08663}
\showURL{%
\tempurl}


\bibitem[Thomas et~al\mbox{.}(2024)]%
        {thomas2024large}
\bibfield{author}{\bibinfo{person}{Paul Thomas}, \bibinfo{person}{Seth
  Spielman}, \bibinfo{person}{Nick Craswell}, {and} \bibinfo{person}{Bhaskar
  Mitra}.} \bibinfo{year}{2024}\natexlab{}.
\newblock \showarticletitle{Large Language Models can Accurately Predict
  Searcher Preferences}. In \bibinfo{booktitle}{\emph{Proceedings of the 47th
  International ACM SIGIR Conference on Research and Development in Information
  Retrieval}} (Washington DC, USA) \emph{(\bibinfo{series}{SIGIR '24})}.
  \bibinfo{publisher}{Association for Computing Machinery},
  \bibinfo{address}{New York, NY, USA}, \bibinfo{pages}{1930–1940}.
\newblock
\showISBNx{9798400704314}
\href{https://doi.org/10.1145/3626772.3657707}{doi:\nolinkurl{10.1145/3626772.3657707}}


\bibitem[Upadhyay et~al\mbox{.}(2024)]%
        {Upadhyay-2024-UMBRELA}
\bibfield{author}{\bibinfo{person}{Shivani Upadhyay}, \bibinfo{person}{Ronak
  Pradeep}, \bibinfo{person}{Nandan Thakur}, \bibinfo{person}{Nick Craswell},
  {and} \bibinfo{person}{Jimmy Lin}.} \bibinfo{year}{2024}\natexlab{}.
\newblock \bibinfo{title}{UMBRELA: UMbrela is the (Open-Source Reproduction of
  the) Bing RELevance Assessor}.
\newblock
\showeprint[arxiv]{2406.06519}~[cs.IR]
\urldef\tempurl%
\url{https://arxiv.org/abs/2406.06519}
\showURL{%
\tempurl}


\bibitem[Wang et~al\mbox{.}(2022)]%
        {MtCut}
\bibfield{author}{\bibinfo{person}{Dong Wang}, \bibinfo{person}{Jianxin Li},
  \bibinfo{person}{Tianchen Zhu}, \bibinfo{person}{Haoyi Zhou},
  \bibinfo{person}{Qishan Zhu}, \bibinfo{person}{Yuxin Wen}, {and}
  \bibinfo{person}{Hongming Piao}.} \bibinfo{year}{2022}\natexlab{}.
\newblock \showarticletitle{MtCut: A Multi-Task Framework for Ranked List
  Truncation}. In \bibinfo{booktitle}{\emph{Proceedings of the Fifteenth ACM
  International Conference on Web Search and Data Mining}} (Virtual Event, AZ,
  USA) \emph{(\bibinfo{series}{WSDM '22})}. \bibinfo{publisher}{Association for
  Computing Machinery}, \bibinfo{address}{New York, NY, USA},
  \bibinfo{pages}{1054–1062}.
\newblock
\showISBNx{9781450391320}
\href{https://doi.org/10.1145/3488560.3498466}{doi:\nolinkurl{10.1145/3488560.3498466}}


\bibitem[Wang et~al\mbox{.}(2024)]%
        {wang2024textembeddingsweaklysupervisedcontrastive}
\bibfield{author}{\bibinfo{person}{Liang Wang}, \bibinfo{person}{Nan Yang},
  \bibinfo{person}{Xiaolong Huang}, \bibinfo{person}{Binxing Jiao},
  \bibinfo{person}{Linjun Yang}, \bibinfo{person}{Daxin Jiang},
  \bibinfo{person}{Rangan Majumder}, {and} \bibinfo{person}{Furu Wei}.}
  \bibinfo{year}{2024}\natexlab{}.
\newblock \bibinfo{title}{Text Embeddings by Weakly-Supervised Contrastive
  Pre-training}.
\newblock
\showeprint[arxiv]{2212.03533}~[cs.CL]
\urldef\tempurl%
\url{https://arxiv.org/abs/2212.03533}
\showURL{%
\tempurl}


\bibitem[Wang et~al\mbox{.}(2023)]%
        {wang-etal-2023-query2doc}
\bibfield{author}{\bibinfo{person}{Liang Wang}, \bibinfo{person}{Nan Yang},
  {and} \bibinfo{person}{Furu Wei}.} \bibinfo{year}{2023}\natexlab{}.
\newblock \showarticletitle{Query2doc: Query Expansion with Large Language
  Models}. In \bibinfo{booktitle}{\emph{Proceedings of the 2023 Conference on
  Empirical Methods in Natural Language Processing}},
  \bibfield{editor}{\bibinfo{person}{Houda Bouamor}, \bibinfo{person}{Juan
  Pino}, {and} \bibinfo{person}{Kalika Bali}} (Eds.).
  \bibinfo{publisher}{Association for Computational Linguistics},
  \bibinfo{address}{Singapore}, \bibinfo{pages}{9414--9423}.
\newblock
\href{https://doi.org/10.18653/v1/2023.emnlp-main.585}{doi:\nolinkurl{10.18653/v1/2023.emnlp-main.585}}


\bibitem[Wu et~al\mbox{.}(2021)]%
        {wu2021learningtruncaterankedlists}
\bibfield{author}{\bibinfo{person}{Chen Wu}, \bibinfo{person}{Ruqing Zhang},
  \bibinfo{person}{Jiafeng Guo}, \bibinfo{person}{Yixing Fan},
  \bibinfo{person}{Yanyan Lan}, {and} \bibinfo{person}{Xueqi Cheng}.}
  \bibinfo{year}{2021}\natexlab{}.
\newblock \bibinfo{title}{Learning to Truncate Ranked Lists for Information
  Retrieval}.
\newblock
\showeprint[arxiv]{2102.12793}~[cs.IR]
\urldef\tempurl%
\url{https://arxiv.org/abs/2102.12793}
\showURL{%
\tempurl}


\bibitem[Zhang et~al\mbox{.}(2024)]%
        {zhang-etal-2024-two}
\bibfield{author}{\bibinfo{person}{Longhui Zhang}, \bibinfo{person}{Yanzhao
  Zhang}, \bibinfo{person}{Dingkun Long}, \bibinfo{person}{Pengjun Xie},
  \bibinfo{person}{Meishan Zhang}, {and} \bibinfo{person}{Min Zhang}.}
  \bibinfo{year}{2024}\natexlab{}.
\newblock \showarticletitle{A Two-Stage Adaptation of Large Language Models for
  Text Ranking}. In \bibinfo{booktitle}{\emph{Findings of the Association for
  Computational Linguistics: ACL 2024}},
  \bibfield{editor}{\bibinfo{person}{Lun-Wei Ku}, \bibinfo{person}{Andre
  Martins}, {and} \bibinfo{person}{Vivek Srikumar}} (Eds.).
  \bibinfo{publisher}{Association for Computational Linguistics},
  \bibinfo{address}{Bangkok, Thailand}, \bibinfo{pages}{11880--11891}.
\newblock
\href{https://doi.org/10.18653/v1/2024.findings-acl.706}{doi:\nolinkurl{10.18653/v1/2024.findings-acl.706}}


\end{thebibliography}

\end{document}